\documentclass[showpacs,amsmath,amssymb,superscriptaddress,nofootinbib,english]{revtex4}
\pdfoutput=1

\usepackage{graphicx,floatflt}
\usepackage{amsmath}
\usepackage{amssymb}

\usepackage{epsfig}
\usepackage{color}

\usepackage[binary-units=true]{siunitx}
\usepackage{hyperref}

\newcommand{\Mpl}{M_\mathrm{pl}} 

\newcommand{\rr}{\mathrm}
\newcommand{\ns}{n_{\rr s} }

\newcommand{\phic}{\phi_{\rr c}}

\newcommand{\Mpc}{\rr{Mpc}}

\newcommand{\kp}{k_{\rr{p}}}

\newcommand{\be}{\begin{equation}}
\newcommand{\ee}{\end{equation}}
\newcommand{\ba}{\begin{align}}
\newcommand{\ea}{\end{align}}

\begin{document}

\title{Testing kinetically coupled inflation models with CMB distortions}

\author{Rui Dai}
\affiliation{School of Science, Wuhan University of Technology, Wuhan 430070, China}
\author{Yi Zhu} \email{yi.zhu@whut.edu.cn}
\affiliation{School of Science, Wuhan University of Technology, Wuhan 430070, China}
\date{\today}

\begin{abstract}
  Inflation scenarios kinetically coupled with the Einstein tensor have been widely studied. They can be consistent with current observational data. Future experiments on the measurement on CMB distortions will potentially extend information about the scalar spectrum to small scales $1 \Mpc^{-1} \lesssim k \lesssim 2 \times 10^4 \Mpc^{-1}$. By taking the sensitivity of the PIXIE experiment as criterion, we perform a model-oriented analysis of the observational prospects of spectral distortions for kinetically coupled inflation. There are five models that possibly generate a detectable level of distortions, among the 49 single-field inflation models listed in Ref. \cite{Martin2013a}. These models are: hybrid inflation in the valley (VHI), non-canonical K\"{a}hler inflation (NCKI), generalized MSSM inflation (GMSSMI), generalized renormalization point inflation (GRIPI), and running-mass inflation (RMI). Each of these models can satisfy the Planck constraints on spectral tilt and lead to increase power on scales relevant for CMB distortions in a tuned region of their parameter space. The existence of kinetic coupling suppresses the value of the model parameters with mass dimension for VHI, GMSSMI, and GRIPI, such that these three models can be in agreement with their theoretical considerations. However, the tuned regions for all these models fail to satisfy the constraints on tensor modes.
\end{abstract}
\pacs{98.80.Cq}
\maketitle

\section{Introduction}

In the last three decades, observations of the anisotropies of the Cosmic Microwave Background (CMB) and the inhomogeneities of Large Scale Structure (LSS) have provided strong evidence for cosmic inflation. At the same time, they also bring stringent constraints to the sharpness and amplitude of the power spectrum of primordial fluctuations at scales where $k \leq 3 \Mpc^{-1}$. Combining the latest observational data from the PLANCK satellite \cite{Akrami2018,Aghanim2018eyx}, the scalar amplitude is required to be $P_\zeta \simeq 2.10 \times 10^{-9}$, the spectral index $n_s \simeq 0.965$, and the tensor to scalar ratio $r_{0.002} < 0.10$.
Many inflation models have been ruled out or strongly disfavored, such as minimally coupled Higgs inflation, power law inflation with $\phi^4$, and original hybrid model. However, a large number of models still survive. Our knowledge about the details of inflation is limited, especially for its behavior on small scales. There are only upper limits for the primordial curvature perturbations from observations of ultra-compact minihalos \cite{Bringmann2012} and primordial black holes \cite{Josan2009}.

In order to extend the available information on the inflation on small scales, some other signals and observational measurements, such as the 21 cm radiation from the dark age \cite{Furlanetto2009,Pritchard2010,Pritchard2012}, and spectral distortions of the CMB \cite{Chluba:2019kpb,Chluba:2019nxa}, are being seriously considered. Generally speaking, spectral distortions can be produced by energy exchange between CMB photons and matter with several physical mechanisms \cite{Sunyaev:1970er,Danese1982,Chluba2011,
Khatri:2012tv,Sunyaev2013,Chluba2013,Chluba:2016bvg,Chluba:2018cww}, e.g., the decay or annihilation of relic particles, the evaporation of primordial black holes, dissipation of primordial acoustic modes, the thermal Sunyaev-Zeldovich effect after recombination, and dark matter annihilation. Among these processes, distortions caused by the damping of primordial perturbations \cite{Silk:1967kq,Kaiser1983} are relevant to the inflationary epoch. The spectral index distortions from Silk damping are first discovered by Sunyaev and Zeldovich \cite{Sunyaev:1970er}. In 2012 Chluba, Khatri, and Sunyaev \cite{Chluba:2012gq,Khatri2012} improve the formulae for distortions which form the basis for subsequent studies. Observations on distortion can provide information on the sharpness and amplitude of the scalar power spectrum on scales ($1 \Mpc^{-1} \lesssim k \lesssim 2 \times 10^4 \Mpc^{-1} $). The first measurements on CMB distortions were done by the COBE/FIRAS experiment \cite{Fixsen1996}. It not only gave limits on $\mu$-type and $y$-type distortions as $\mu < 9 \times 10^{-5}$ and $y < 1.5 \times 10^{-5}$, but also yielded a limit on the value of the spectral index, $n_s < 1.6$ \cite{Hu:1994bz}. In 2011, the Primordial Inflation Explorer (PIXIE) experiment \cite{Kogut2011,Kogut:2019vqh} was proposed as part of NASA's MIDEX program. It is designed to improve the limit on CMB distortions by about three orders of magnitude. A more ambitious proposal is the Polarized Radiation Imaging and Spectroscopy Mission (PRISM) \cite{Andre2013,Andre2013a,Kogut:2019vqh} with a sensitivity which is about ten times better than that of PIXIE. Although these two missions are not settled yet, one can still discuss the probabilities in the future experiments. A number of authors have forecasted constraints on inflation using a wide range of models or methods \cite{Chluba2012,Pajer:2012vz,Khatri2013,Chluba:2013dna,Emami:2015xqa,Khatri:2015tla, Shiraishi:2015lma,Dimastrogiovanni:2015wvk,Cabass:2016ldu,Chluba:2016bvg,Shiraishi:2016hjd, Chluba:2016aln,Nakama:2017ohe,Haga:2018pdl,Remazeilles:2018kqd,Chluba:2019kpb}. There are also many works that estimate the level of CMB distortions relevant to inflation scenarios \cite{Ota:2014hha,Chluba:2014qia,Ota:2014iva,Clesse:2014pna, Clesse:2015wea,Enqvist:2015njy,Cho:2017zkj,Kainulainen:2017gqq,Bae:2017tll}. In Ref. \cite{Clesse:2014pna}, the authors have performed a model-oriented analysis of the observational prospects for inflation, and find that few models can lead to detectable signals for the PIXIE experiment.

On the theoretical side, as the most general second-order scalar-tensor theory in four dimensions, Horndeski theory, have recently been investigated with many cosmological consequences \cite{Kobayashi:2019hrl,Charmousis:2011bf,Darabi:2013caa,Sushkov:2012za, Gumjudpai:2015vio,Skugoreva:2013ooa}. In Horndeski theory \cite{Horndeski:1974wa,Kobayashi:2019hrl}, the curvature tensors couple with a scalar field in 6 types in the case of four derivatives. In Ref. \cite{Capozziello:1999xt}, the authors show that leaving only $R g^{\mu \nu} \partial_\mu \phi \partial_\nu \phi$ and $R^{\mu \nu}  \partial_\mu \phi \partial_\nu \phi$ terms is enough to preserve generality, and all other coupling terms are not necessary. Generally, with non-minimal derivative coupling, the order of field equation is higher than two. However, there is a ghost-free combination of these coupling terms, where the Einstein tensor $G^{\mu \nu} = R^{\mu \nu} - \frac{1}{2} g^{\mu \nu} R$ coupled with the kinetic term of the scalar field (referred as kinetic coupling). A cosmological model with this coupling term can explain both a quasi-de Sitter phase in the early universe and an exit without any fine-tuned potential \cite{Sushkov:2009hk}. For a scalar field theory without kinetic energy, the non-minimal coupled term $G^{\mu \nu}  \partial_\mu \phi \partial_\nu \phi$ can behave like dark matter, whereas the scalar potential can play a role as dark energy \cite{Gao:2010vr}. For inflation scenarios, the non-minimal derivative coupling provides an extra friction to the rolling processes of the inflation. In this circumstance, the slow-roll conditions are easier to satisfy and the inflation process can last longer than that of the minimal coupled case. Many inflation models including those that were seemingly ruled out by observations have been re-investigated \cite{Germani:2010gm,Yang:2015pga,Dalianis:2015aba,Myung:2015tga,Qiu:2015aha}. Curvaton scenarios and reheating processes have been reconsidered as well \cite{Herrera:2016fdo,Dalianis:2016wpu,Feng:2014tka}. With kinetic coupling, these models can easily satisfy the constraints from CMB anisotropy observations. Therefore, it is worthwhile to investigate the non-minimally coupled inflation scenarios on the small-scale behavior of the scalar power spectrum, and weigh the possibility of inducing observational signals for future experiments.

The paper is organized as follows. In Sec. \ref{sec:dynamics}, we briefly review the dynamics of non-minimal derivative-coupled inflation and the induced scalar power spectrum of perturbations. Sec. \ref{sec:crit} is dedicated to develop criteria for observable spectral distortion signals with the sensitivity of a PIXIE-like experiment. In Secs. \ref{sec:VHI} to \ref{sec:RMI}, we study the properties of CMB distortions for single field models in the high friction limit. Finally, we summarize our results and discuss the implications for future CMB distortion experiments.

\section{Non-minimal Derivative-coupled inflation} \label{sec:dynamics}
We consider a scalar field theory, which is kinetically coupled with the Einstein tensor. The action is given by
\begin{equation} \label{eq:action}
S = \frac{1}{2} \int d^4 x \sqrt{-g} \left[ \Mpl^2 R - g^{\mu \nu} \partial_\mu \phi \partial_\nu \phi + \frac{1}{M^2} G^{\mu \nu}  \partial_\mu \phi \partial_\nu \phi - 2 V(\phi) \right]
\end{equation}
where $\phi$ is the scalar field, $V(\phi)$ is its potential, $\Mpl$ is the reduced Planck mass, and $M$ is the coupling constant with dimension of mass. For a spatially flat maximally symmetric spacetime, where $ds^2 = - dt^2 + a^2(t) d\Omega$, one obtains the Friedmann equation as
\begin{equation}
H^2 = \left( \frac{\dot{a}}{a} \right)^2 = \frac{1}{3 \Mpl^2} \left[ \frac{\dot{\phi}^2}{2} \left( 1 + 9 \frac{H^2}{M^2} \right) +V(\phi) \right],
\end{equation}
 and the Klein-Gordon equation (equation of motion for the scalar field) as
\begin{equation}
\frac{d}{dt} \left[ a^3 \dot{\phi} \left( 1 + 9 \frac{H^2}{M^2} \right)  \right] = - a^3 \frac{d V}{d \phi}.
\end{equation}
where $H \equiv \left( \frac{\dot{a}}{a} \right)$ is the Hubble expansion rate, and a dot denotes a derivative with respect to the cosmic time $t$.

The non-minimal derivative coupling plays a role as a friction force, which slows down the rolling speed of the scalar field. In order to look clearly into the effect of the derivative coupling, we consider the high friction limit where $M^2 \ll H^2$.  Inflation can be approximately described by a slow-roll process. The first and second slow parameters are defined respectively by
\be \label{eq:epsilon}
\epsilon = \frac{\Mpl^2}{2} \left( \frac{V_\phi}{V} \right)^2 \frac{M^2}{3 H^2}, \quad \eta = \Mpl^2 \frac{V_{\phi \phi}}{V} \frac{M^2}{3H^2}
\ee
with $V_\phi \equiv d V/d \phi$ and $V_{\phi \phi} \equiv d^2 V / d \phi^2$. The slow roll parameters are suppressed by a factor of $M^2/3H^2$. It is therefore much easier to trigger the accelerated expansion of the spacetime. In the slow roll approximation, the number of e-folds can be calculated by
\be
N(\phi_*) = \int_{\phi_*}^{\phi_e} \frac{H}{\dot{\phi}} d \phi = \int_{\phi_*}^{\phi_e} \frac{1}{\sqrt{2 \epsilon}} \sqrt{\frac{3H^2}{M^2}} \frac{d \phi}{\Mpl},
\ee
where $\phi_*$ and $\phi_e$  are the field values at horizon exit and at the end of inflation, respectively. To first order in the slow-roll parameters, the scalar power spectrum is given by
\be
\mathcal{P}_\zeta (k) = \frac{H_k^2}{8 \pi^2 \Mpl^2 \epsilon_k} \left[ 1 + \left( \frac{1}{3} - 8C \right) \epsilon_k - \left( \frac{2}{3} -2C \right) \eta_k \right],
\ee
and the spectral index reads
\be \label{eq:index}
n_s (k) = 1 - 8 \epsilon_k + 2 \eta_k,
\ee
where $C \equiv \gamma_E + \ln 2 -2 \simeq -0.7296$ and $\gamma_E$ is the Euler constant. The quantities with subscript $k$ are evaluated at the time $t_k$ of horizon exit of the corresponding mode, i.e. when $k = a(t_k) H(t_k)$. The tensor spectrum amplitude and index are given by
\be
\mathcal{P}_T (k) = \frac{2 H_k^2}{\pi^2 \Mpl^2} \left[ 1 - \left( \frac{8}{3} - 2 C\right) \epsilon_k \right],
\ee
\be
n_T (k)= -2 \epsilon_k.
\ee
Thus one obtain the tensor-to-scalar ratio
\be \label{eq:ratio}
r(k) =\frac{P_T}{P_\zeta} = 16 \epsilon_k.
\ee
The standard inflationary consistency relation $n_T = -r/8$ is still valid. In the squeezed limit, the non-Gaussian parameter reads \cite{Maldacena:2002vr,Germani:2011ua}
\be
\frac{12}{5} f_{NL}^{local} = \frac{B_\zeta (k_1, k_2 \rightarrow k_1, k_3 \rightarrow 0)}{P_\zeta (k_1) P_\zeta (k_3)} = 1 - n_s,
\ee
where $B_\zeta (k_1, k_2, k_3)$ is the scalar bispectrum. This equation is called the consistency relation for the three-point function \cite{Creminelli:2004yq}. In Ref. \cite{Gao:2011qe,DeFelice:2011uc}, the authors showed that the parameter $f_{NL}^{equil} \simeq \mathcal{O}(\epsilon, \eta)$ for the case of non-minimal derivative coupling, since the scalar propagation speed is close to the speed of light. Therefore, for kinetically coupled single field models, one do not need to consider the constraints on non-Gaussianity which are $f_{NL}^{local} = -0.9 \pm 5.1$, and $f_{NL}^{equil} = -26 \pm 47$ from the recent Planck observation \cite{Akrami:2019izv}.

The latest CMB anisotropies measurements give stringent constraints on the amplitude and spectral index of scalar power spectrum. We apply the constraints from PLANCK 2018 \cite{Aghanim2018eyx,Akrami2018}. They give
\be
\label{values:fiducial}
\mathcal P_\zeta(\kp) = 2.10 \pm 0.03 \times 10^{-9} , \hspace{10mm} \ns(\kp) = 0.965 \pm 0.004~,
\ee
at pivot scale $k_p = 0.05 \Mpc^{-1}$, which exits the horizon about 60 e-folds before the end of inflation. For primordial tensor fluctuations, the joint analysis of PLANCK 2018 and BICEP2-Keck Array data gives a limit on the scalar-to-tensor ratio \cite{Array:2015xqh,Aghanim2018eyx,Akrami2018}
\be
r_{0.002} < 0.065
\ee
at $95\%$ confidence level (CL).
Since the discrepancies for the scalar spectral index and amplitude between recent PLANCK 2018 data \cite{Aghanim2018eyx,Akrami2018} and earlier PLANCK 2013 data \cite{Ade2013} are very small, we can make a direct comparison with the results from the minimally coupled models in Ref. \cite{Clesse:2014pna}.

\section{Criteria for observable CMB distortions}  \label{sec:crit}
When scalar fluctuations re-enter into the horizon, an inevitable $y$-type distortion of the CMB spectrum is caused by dissipating acoustic waves in the Silk-damping tail. In the early universe, it is converted to a $\mu$-distortion or an intermediate $i$-distortion by the thermalization process and Compton scattering. The first measurement of $\mu$-type spectral distortion has been done by COBE/FIRAS. It gives an upper limit for the spectral index $n_s \leq 1.6$ \cite{Hu:1994bz}. A recent model independent analysis of CMB spectral distortions from Ref. \cite{Chluba2012} have updated the limit to be $n_s \leq 3$ at the scale $k_d =42 \Mpc^{-1}$. As for future experiments, the designed sensitivity of PIXIE allows a detection of spectral distortions characterized by \cite{Kogut2011}
\be
\mu = 5 \times 10^{-8} \quad {\rr and} \quad y = 1 \times 10^{-8}  \;\; {\rr at} \;\; 5 \sigma,
\ee
which is a thousand times improved from COBE/FIRAS. However, the damping signal is close to the detection limit of the PIXIE experiment. Using a Bayesian Markov-Chain-Monte-Carlo study with fiducial values $\mathcal{P}_\zeta^{fid} (k_d = 42\Mpc^{-1} = 1.68 \times 10^{-9})$, $n_s = 0.96$ and $n_{run}=0$, Ref. \cite{Clesse:2014pna} suggests one take
\be
\mathcal{P}_\zeta (k_d = 42\Mpc^{-1}) = 2.8 \times 10^{-9},
\ee
as the lower limit of a $2 \sigma$ detection. Therefore, observable CMB distortions must come from inflation models which satisfy PLANCK constraints at CMB anisotropy scales and give rise to an enhanced power on small scales. In addition, we do not need to consider y-distortions from the dispersion of acoustic modes, since it cannot be distinguished from the dominant contribution of the thermal Sunyaev-Zeldovich effect.

For further investigation on inflation models, we use the criteria from Ref. \cite{Clesse:2014pna} on the scalar spectrum for potential interesting inflation models, with minor modifications:
\begin{enumerate}
  \item There exists a locally blue tilted spectrum where $n_s > 1$. It requires that the slow-roll parameters satisfy $\eta > 4\epsilon >0$ at some scale.
  \item The scale of the blue spectrum must be smaller than that of the CMB anisotropy observations, i.e. a phase with $n_s > 1$ must follow the phase with $n_s < 1$. In order to ensure the number of e-folds never diverges, we also require that the first slow-roll parameter does not vanish.
  \item We also demand inflation models generate a power spectrum with $n_s = 0.965 \pm 0.004$ at the PLANCK observation pivot scale  $k_p = 0.05 \Mpc^{-1}$, and $n_s >1$ at the pivot scale of CMB distortions $k_d = 42 \Mpc^{-1}$.
  \item The constraints on tensor modes must be satisfied, i.e. the tensor-to-scalar ratio $r_{0.002} < 0.065$.
\end{enumerate}
We add kinetic coupling \eqref{eq:action} to all 49 signal field inflation models listed in Ref. \cite{Martin2013a}. By applying criteria 1 and 2, large classes of models are eliminated. Only five models survive. They are hybrid inflation in a valley, generalized MSSM inflation, generalized renormalizable inflection point inflation, non-canonical K\"{a}hler inflation, and running-mass inflation. This result is the same as that of the minimally coupled cases in Ref. \cite{Clesse:2014pna}. The reason is that the derivative coupling gives a suppression factor $\frac{M^2}{3 H^2}$ to slow-roll parameters, but do not change their sign. The sign of the spectral index $n_s -1$ is therefore preserved. For applying the third criterion, we numerically integrate the slow-roll dynamics and figure out potentially interesting regions in $n_s$ contour plot for each of these models. The overlap between the potentially interesting region and the area allowed by tensor mode constraints is the parameter region that may give rise to observable CMB distortions for a PIXIE-like experiment. In the following sections, we explore the parameter space for these five models in detail.

\section{Hybrid inflation in the Valley (VHI)} \label{sec:VHI}
\begin{figure}
  \centering
  \includegraphics[scale=0.5]{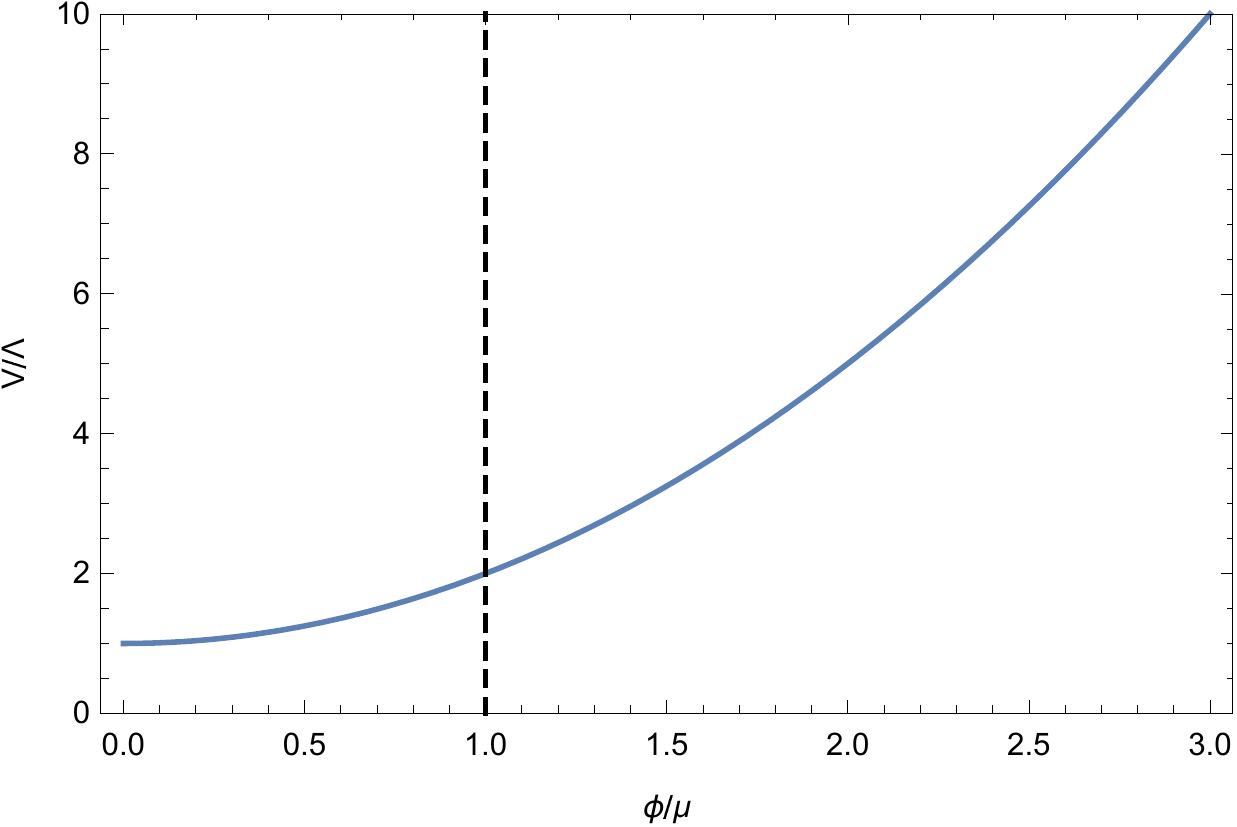}~~~~\includegraphics[scale=0.5]{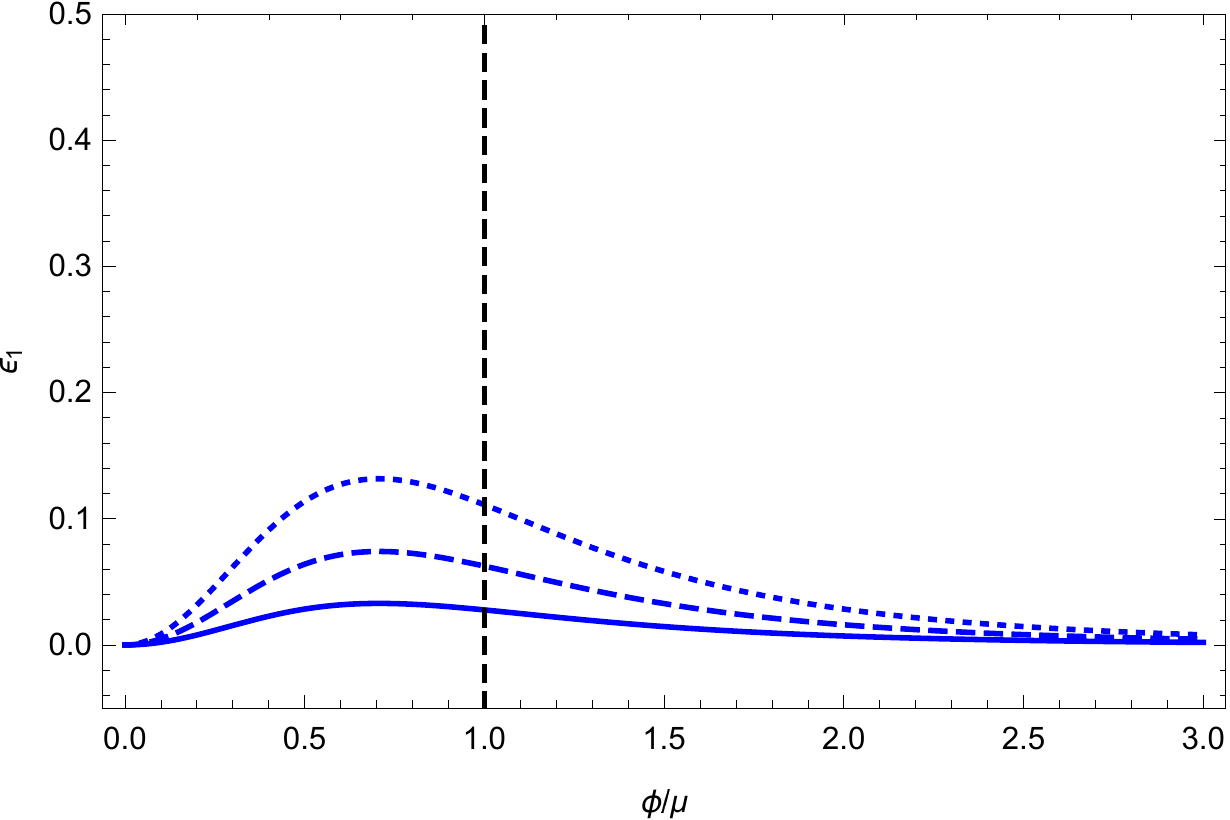}
  \includegraphics[scale=0.5]{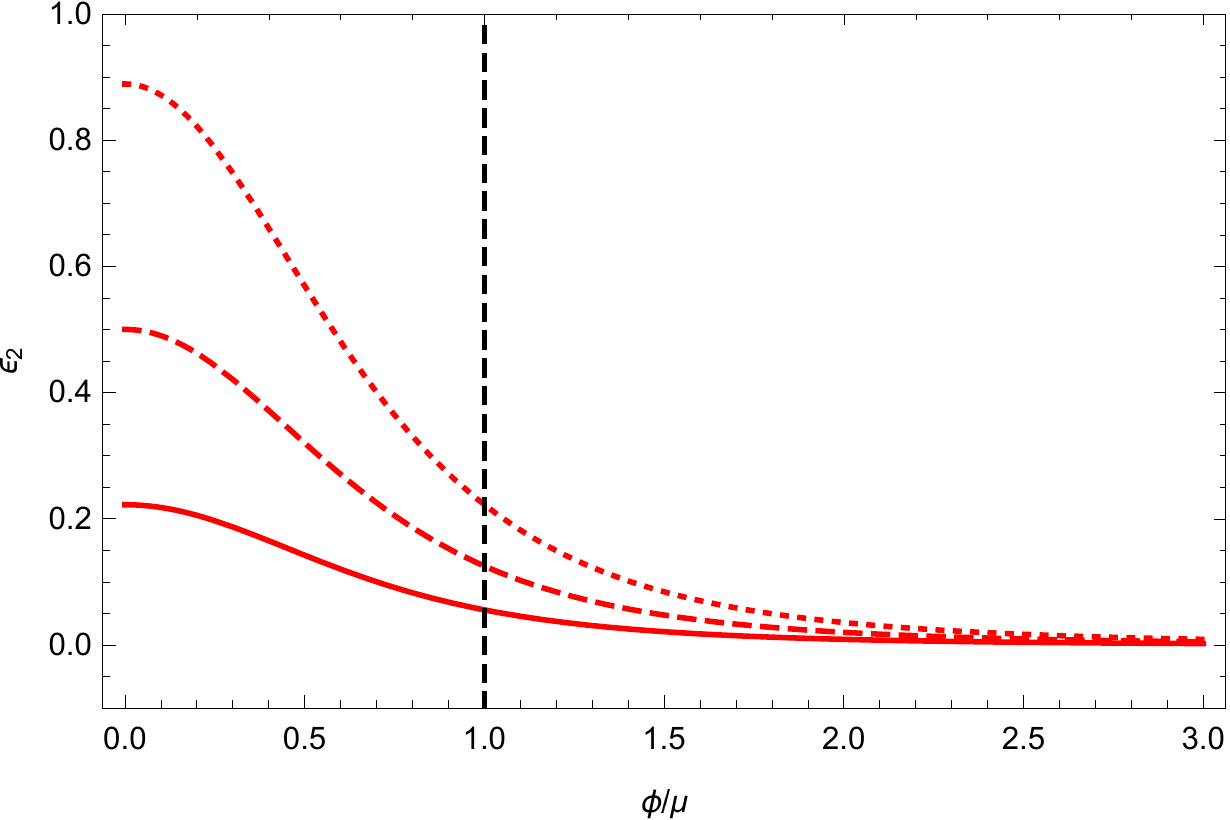}~~~~\includegraphics[scale=0.5]{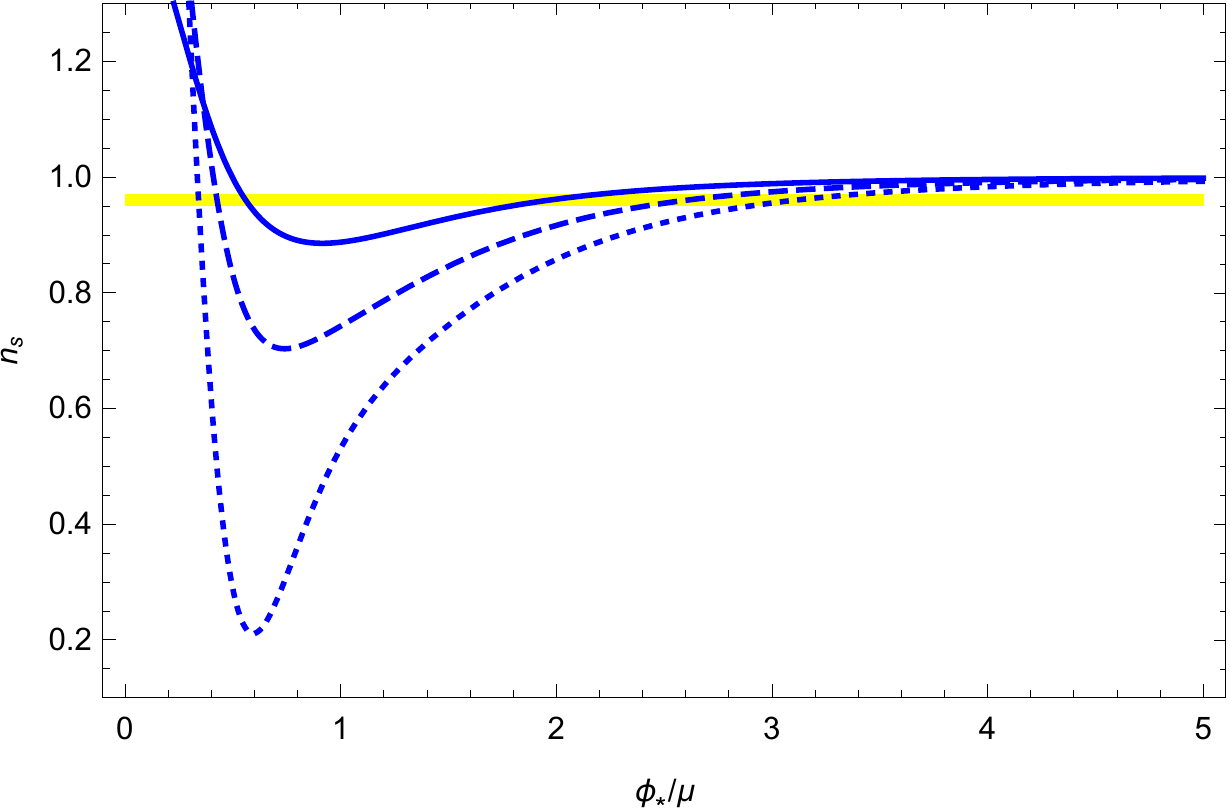}\\
  \caption{Top left: original hybrid potential.  Top right and bottom left: first and second slow-roll parameters $\epsilon$ and $\eta$. Bottom right: scalar spectral index $\ns$ as a function of $\phi_k$. The horizontal band corresponds to the PLANCK $95 \%$ CL constraints. The parameter values in the diagrams for $\epsilon_{1,2}$ and $n_s$ are $\mu_{\rr{VHI}} = 0.015 \Mpl$ (dotted), $\mu_{\rr{VHI}} = 0.02 \Mpl$ (dashed) and $\mu_{\rr{VHI}} = 0.03 \Mpl$ (solid). We are interested in the region close to $\phi_k /\mu_{\rr{VHI}} = 1$, where the power spectrum of curvature perturbations is red-tilted, but which becomes blue-tilted a few e-folds later (i.e. at smaller values of $\phi_k$.)}\label{fig:pot_HVI}
\end{figure}

In the original hybrid inflation model \cite{Linde1994}, there are two scalar fields. One is the inflaton, the other is the waterfall field. Inflation takes place when the inflation rolls slowly down a potential valley. As the inflation reaches the critical point, it triggers a tachyonic waterfall phase, where the waterfall field rolls rapidly, then inflation ends. The dynamics of the inflation process can be described by an effective single field potential
\be
V(\phi)=\Lambda\left(1+\frac{\phi^2}{\mu^2_{\rr VHI}}\right).
\ee
The potential is plotted in the top left panel of Fig. \ref{fig:pot_HVI}. The inflaton rolls from the right side towards the decreasing field values. The value of the inflaton field at the end of inflation $\phi_{end}$ is considered a free parameter. In the high friction limit, the first and second slow-roll parameters are given by
\begin{eqnarray}
  \epsilon &=&\frac{M^2}{3H^2}\frac{2 x^2 \Mpl^2}{\mu^2 (1+x^2)^2} = \frac{2 \times 10^{-4} x^2 \Mpl^2}{\mu^2 (1+x^2)^3}, \label{eq:eps1_HVI}\\
  \eta &=& \frac{M^2}{3H^2}\frac{2 \Mpl^2}{\mu^2 (x^2+1)} = \frac{2 \times 10^{-4} \Mpl^2}{\mu^2 (x^2+1)^2}, \label{eq:eps2_HVI}
\end{eqnarray}
where $x = \phi / \mu_{\rr VHI}$. In these equations, the slow-roll approximation $3\Mpl^2 H^2 = V$ has been applied. We take the non-minimal derivative coupling constant $M = 0.01 \sqrt{\Lambda} / \Mpl$. Slow roll parameters are also presented in Fig. \ref{fig:pot_HVI}. The first criterion from Sec. \ref{sec:crit}, $\eta > 4\epsilon >0$ can be satisfied when the field value $x$ is small enough. In this situation, the Hubble parameter $H^2$ is of the same order as $\Lambda / \Mpl^2$, such that $M^2 \ll H^2$. It is in the high friction limit where the effects of the kinetic coupling are magnified. In addition, the spectral index as a function of $\phi_*$ is also plotted. From Fig. \ref{fig:pot_HVI}, one notes that the phase of the red spectrum is earlier than the phase of the blue spectrum during the inflation epoch. The non-minimal derivative-coupled hybrid inflation may therefore satisfy simultaneously PLANCK's constraints and generate an observable CMB distortion signal.

\begin{figure}
  \centering
  \includegraphics[width=12cm]{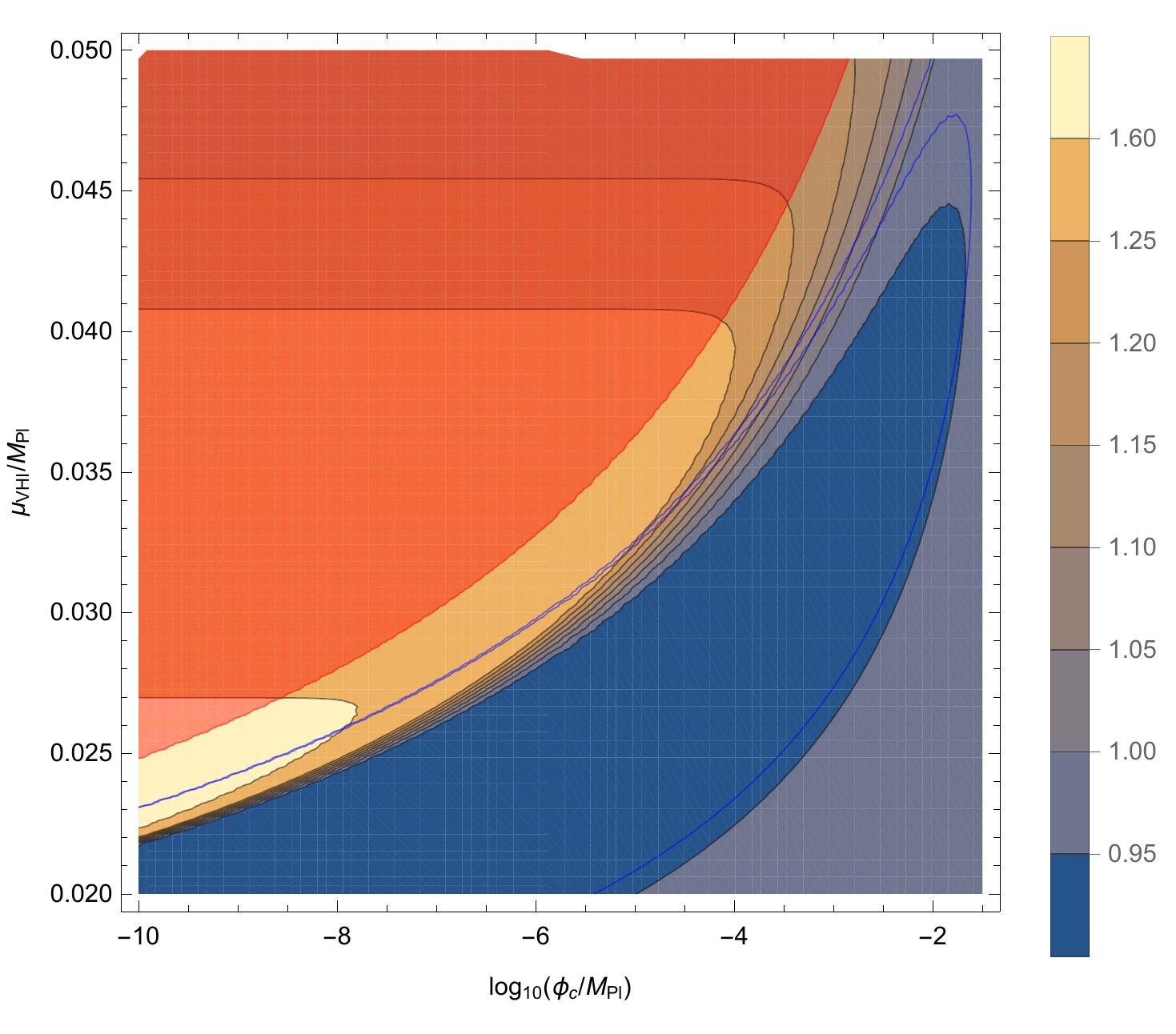}\\
  \caption{Contours of spectral index values for VHI with the non-minimal derivative coupling $M = 0.01 \sqrt{\Lambda} / \Mpl$ at the pivot scale of CMB distortions $k_d=42 \Mpc^{-1}$. The area between the solid blue lines is consistent with PLANCK constraints on scalar spectrum at $95\%$ CL. The red region where the corresponding tensor-to-scalar ratio $r_{0.002} < 0.65$ is in consistent with the constraints on tensor modes.}\label{fig:ns_HVI}
\end{figure}

\begin{figure}
  \centering
  \includegraphics[width=12cm]{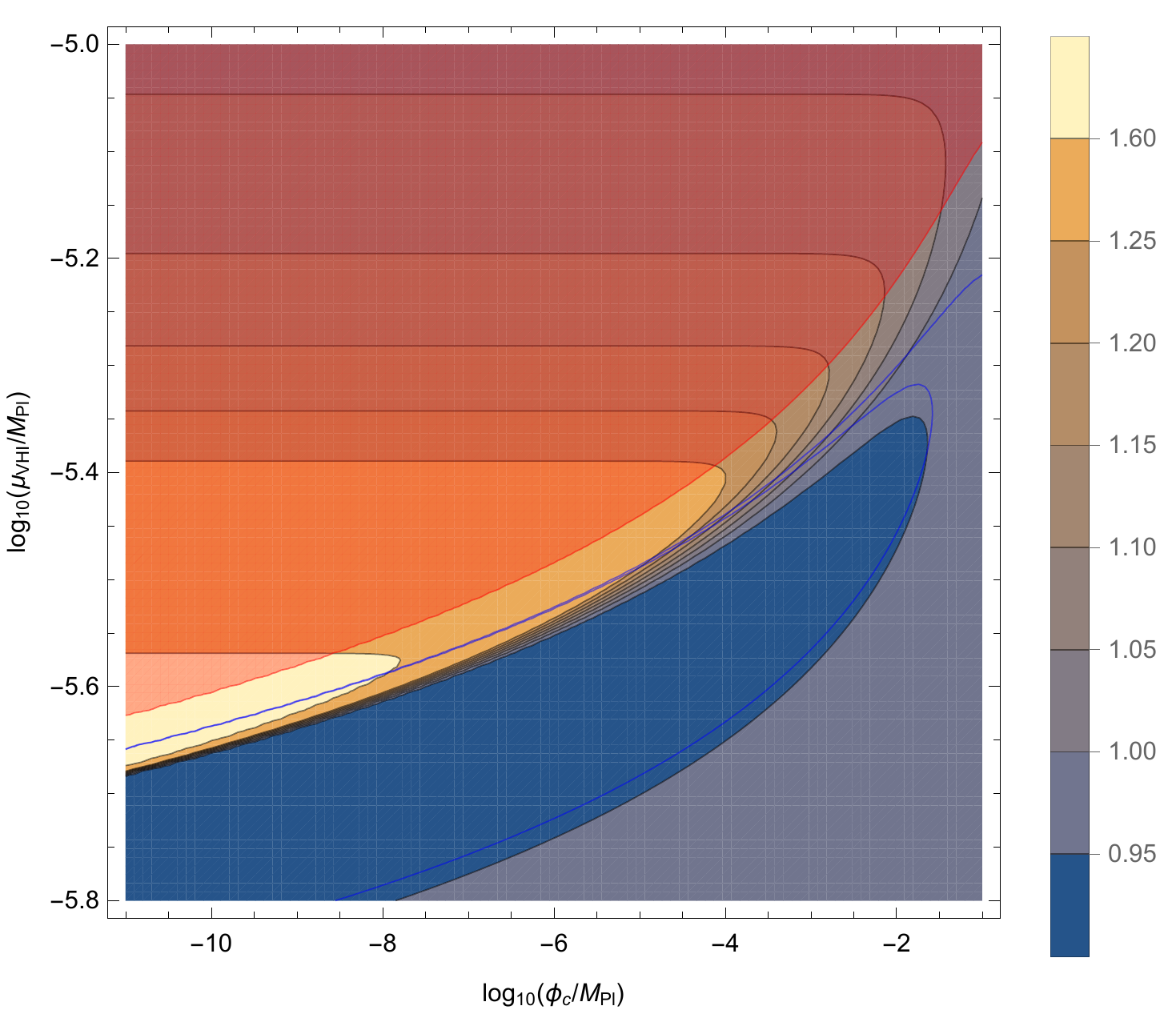}\\
  \caption{Same as in Fig. \ref{fig:ns_HVI} with the coupling constant $M = 10^{-6} \sqrt{\Lambda} / \Mpl$.}\label{fig:ns_HVI_str}
\end{figure}

To check the third criterion, we integrate the Klein-Gordon equation numerically in slow-roll inflation, and draw a contour plot for the spectral index value in the parameter space of $\mu_{VHI}$ and $\phi_{end}$ in Fig. \ref{fig:ns_HVI}. The blue solid lines represent the PLANCK constraints at $95\%$ CL on scalar spectrum. The overlap of the region between the blue lines and those of $n_s >1$ is the region of interest. We note that the value of parameter $\mu_{\rr{VHI}}$ of this potentially interesting region is suppressed to be sub-Planckian. Then we evaluate the tensor-to-scalar ratio at the scale $k_r = 0.002 \Mpc^{-1}$. The red region in Fig. \ref{fig:ns_HVI} is the allowed area for tensor mode constraints. It does not overlap with the potential interest region for scalar modes. Therefore, in the case of $M = 0.01 \sqrt{\Lambda} / \Mpl$, the VHI model cannot satisfy the observation constraints and lead to a detectable CMB distortion signal simultaneously. For an extremely strong coupling case where $M = 10^{-6} \sqrt{\Lambda} / \Mpl$, the conclusion is the same, as shown in Fig. \ref{fig:ns_HVI_str}. The reason is that values of model parameter $\mu_{VHI}$ in the potentially interesting area change proportionally with the coupling constant $M$. According to Eq. \eqref{eq:eps1_HVI} and \eqref{eq:eps2_HVI}, the slow roll parameters are therefore preserved, such that behaviors of $n_s$ and $r_{0.002}$ are similar for different values of $M$.

\section{Non-canonical K\"ahler Inflation (NCKI)}
\begin{figure}
  \centering
  \includegraphics[scale=0.5]{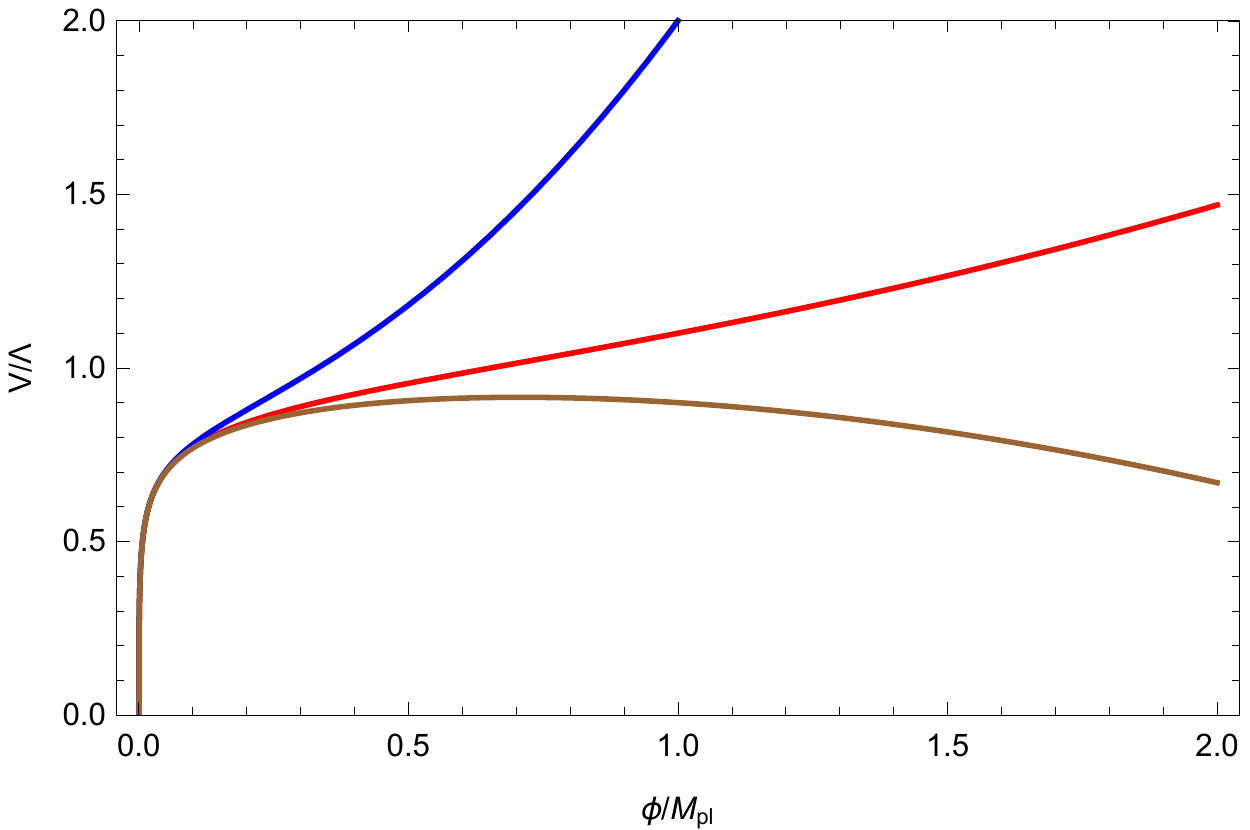}~~\includegraphics[scale=0.5]{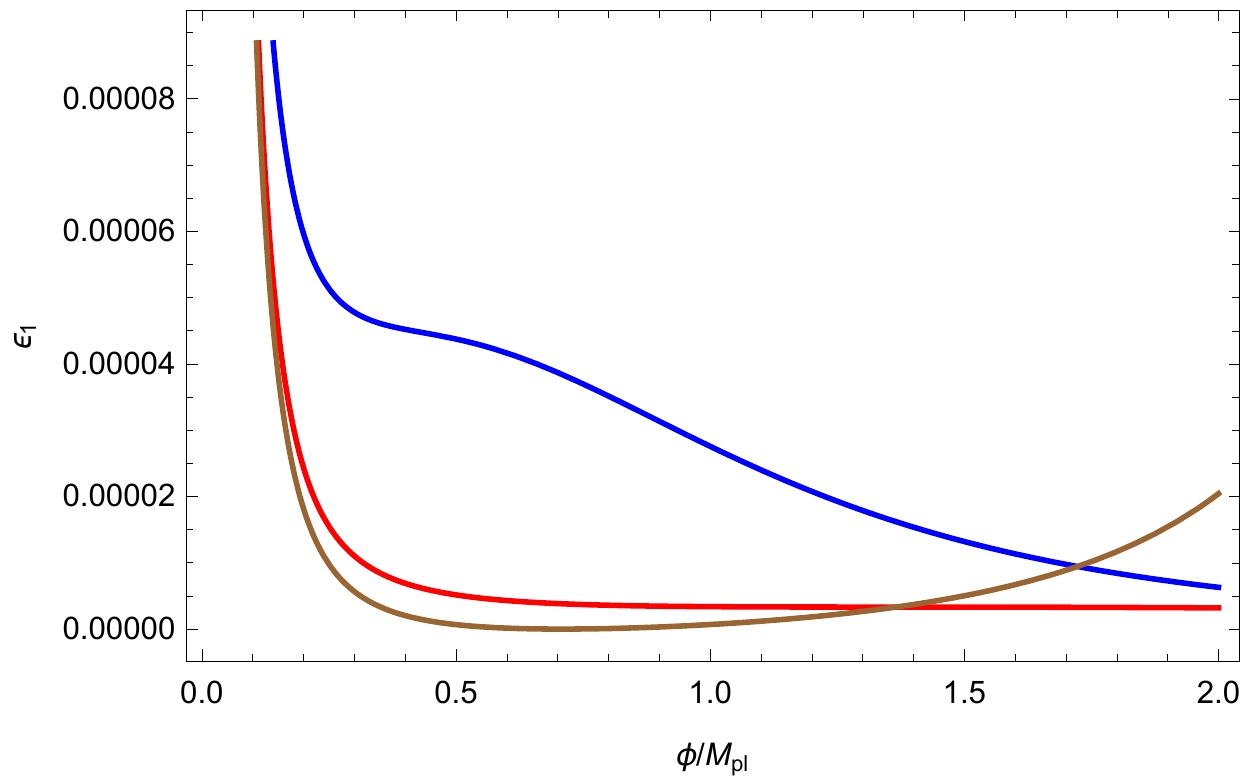}
  \includegraphics[scale=0.5]{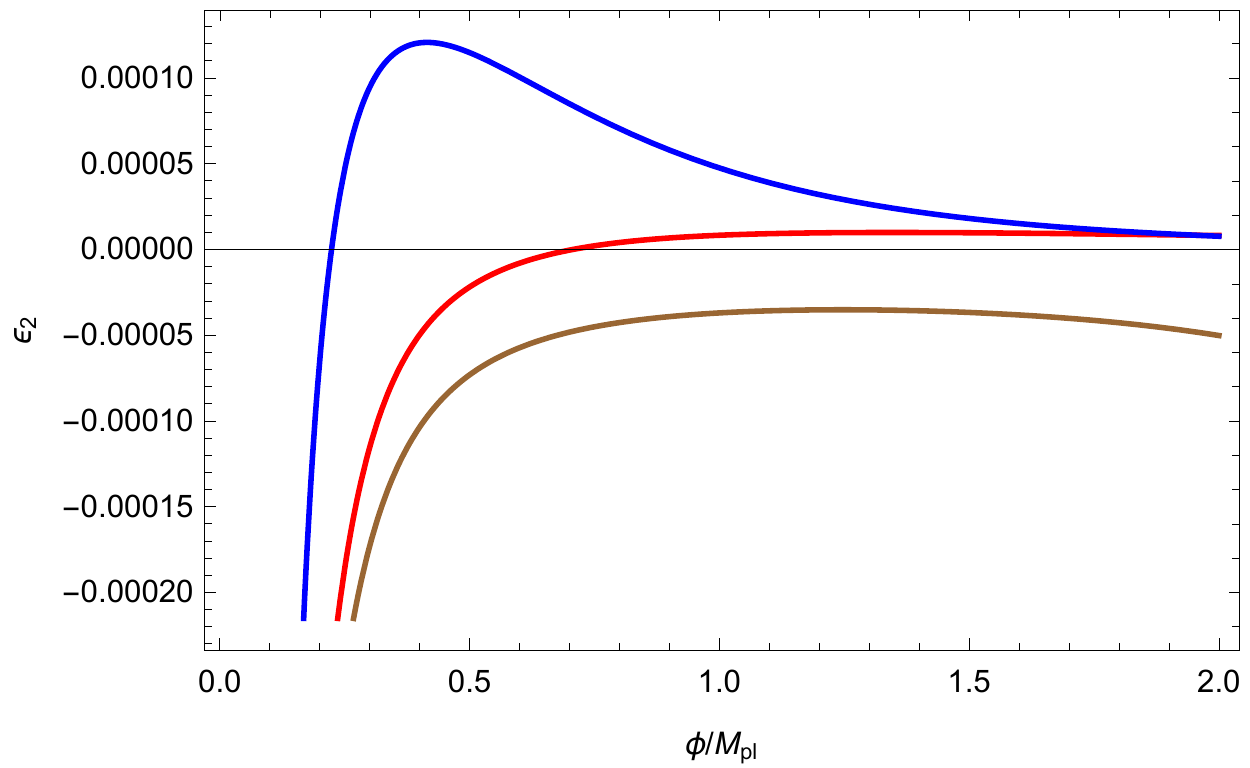}~~\includegraphics[scale=0.5]{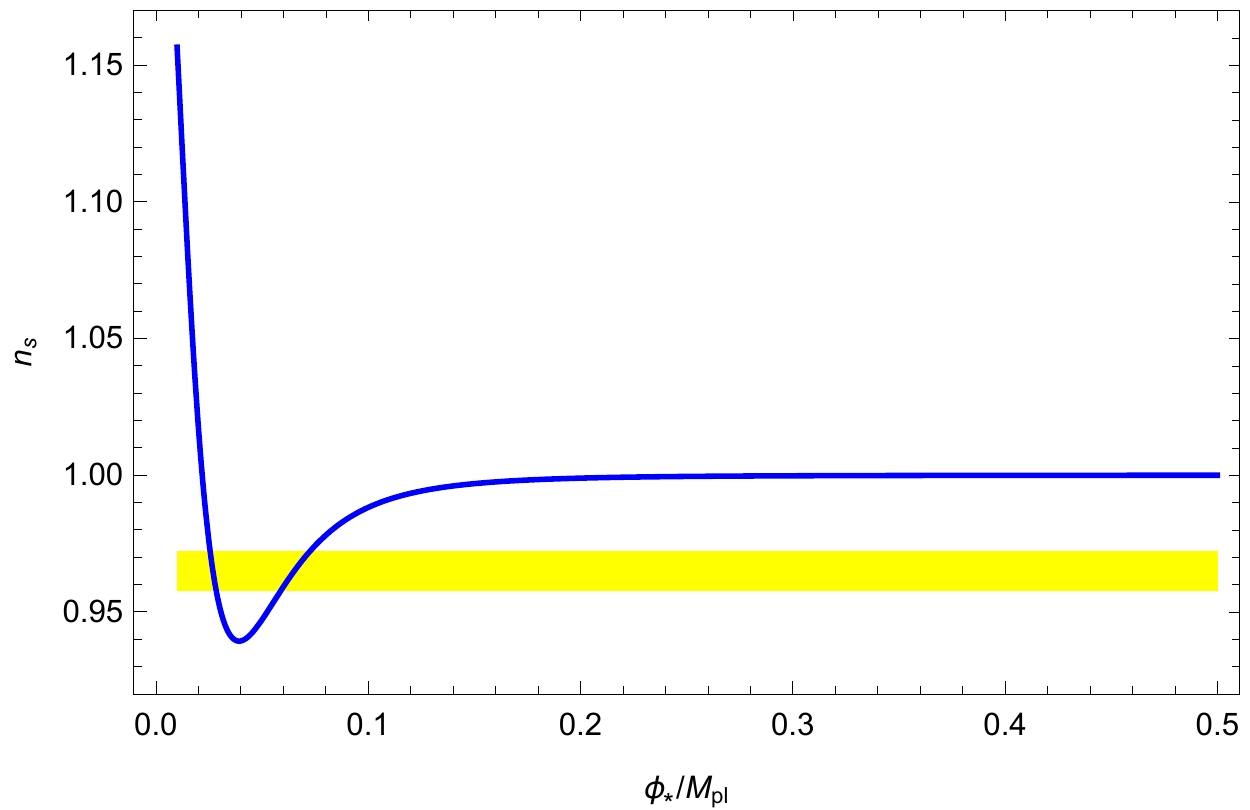}\\
  \caption{Top left: Potential of the NCKI model.Top right and bottom left: first and second slow-roll parameter $\epsilon,\eta$. The parameter sets for these three diagrams are: $\alpha=0.1,\beta=1$(blue), $\alpha=0.1,\beta=0.1$(red), and $\alpha=0.1,\beta=-0.1$(brown). Bottom right: spectral index value as a function of $\phi_*$ for $\alpha=10^{-7}$ and $\beta=600$.}\label{fig:pot_NCKI}
\end{figure}

Non-canonical K\"ahler inflation is derived from the usual hybrid model by taking additional corrections from higher order operators in supergravity. The potential is given by
\be \label{eq:pot_NCKI}
V = \Lambda (1 + \alpha \ln x + \beta x^2),
\ee
where $x=\phi/ \Mpl$, and $\alpha$ is a positive dimensionless parameter. The value of the free parameter $\beta$ can be either positive and negative. The potential is plotted in the top-left panel of Fig. \ref{fig:pot_NCKI} with different choices of parameters. With non-minimal derivative coupling the first and second slow-roll parameters are given by
\begin{eqnarray}
\epsilon &=& \Mpl^2 \frac{M^2(\frac{\alpha}{x}+2x \beta)^2}{2\Lambda(1+x^2{\beta}+{\alpha}{\ln[x]})^2}, \\
\eta &=& \Mpl^2 \frac{M^2(-\frac{\alpha}{x^2}+2\beta)}{\Lambda(1+x^2\beta+\alpha\ln[x])^2},
\end{eqnarray}
Compared with the minimal coupled case, one notes that variations of slow-roll parameters are strongly suppressed. Besides, when $\beta < 0$ the field potential develops a maximum hill. Inflation can occur by rolling down the field from both sides of the potential hill. However, the value of $\eta > 0$ is always negative, such that the $\beta<0$ case is therefore not of interest, which is similar to the minimal coupled NCKI.

\begin{figure}
  \centering
  \includegraphics[width=12cm]{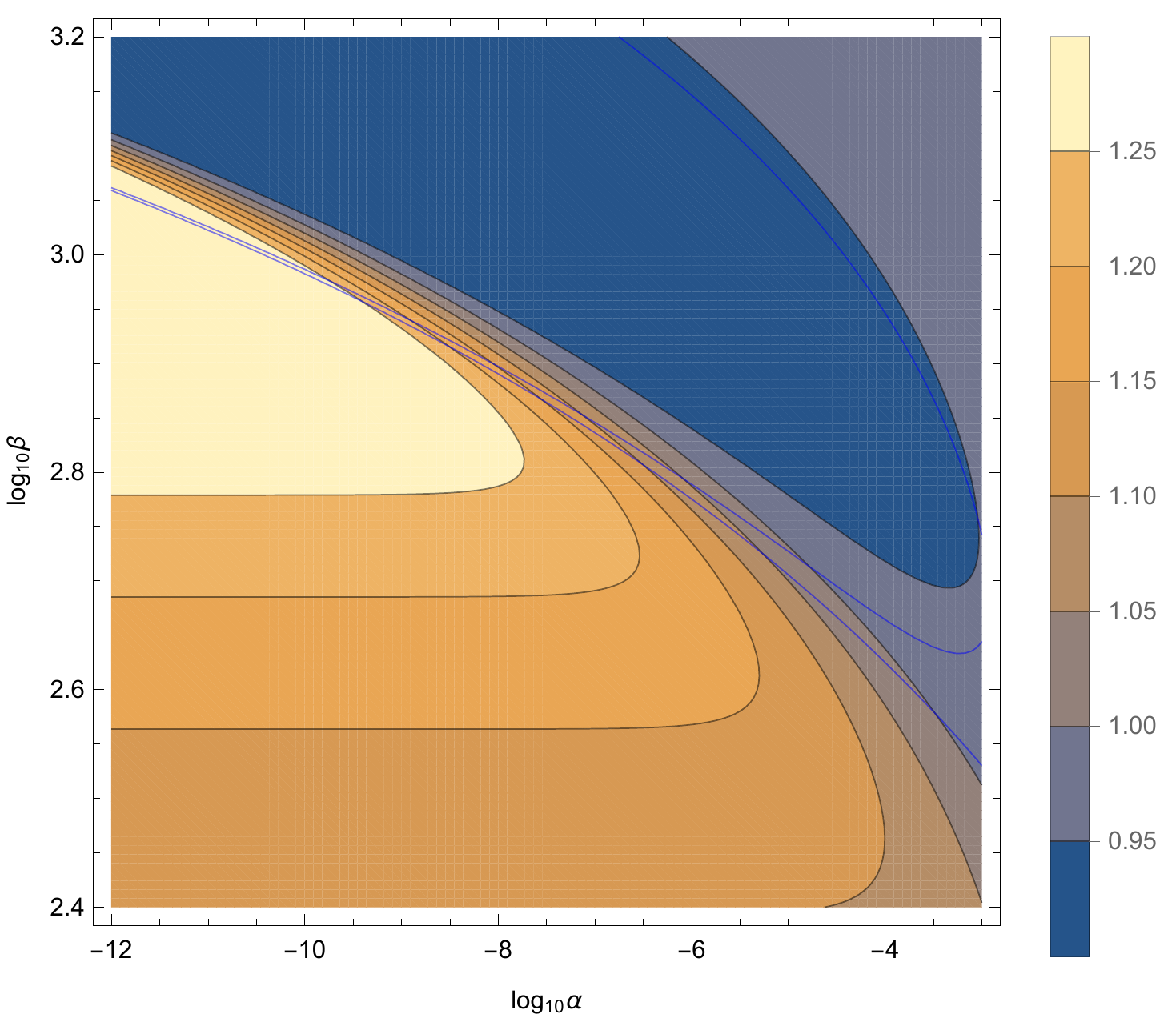}\\
  \caption{Contours of spectral index values for NCKI.The area between the solid blue lines is consistent with the spectral index at $k_p=0.05 \Mpc^{-1}$ at $95\%$ CL.}\label{fig:ns_NCKI}
\end{figure}

For $\beta>0$, we look into the high-friction limit by setting $M = 0.01 \sqrt{\Lambda} / \Mpl$, and draw a contour diagram of scalar index at the distortion pivot scale $k = 42 \Mpc^{-1}$ in the plane of $\alpha$ and $\beta$, as shown in Fig. \ref{fig:ns_NCKI}. The PLANCK-allowed region is between the two blue lines. From this figure, we find that, to satisfy PLANCK constraints on scalar spectrum and generate a detectable distortion signal simultaneously, the parameter $\alpha$ must be much smaller than $\beta$. In this situation, the potential \eqref{eq:pot_NCKI} has reduced to that of the VHI model. Therefore, this model does not give rise to any detectable distortion of the CMB spectrum either.

\section{Generalized MSSM Inflation (GMSSMI)}
\begin{figure}
  \centering
  \includegraphics[scale=0.5]{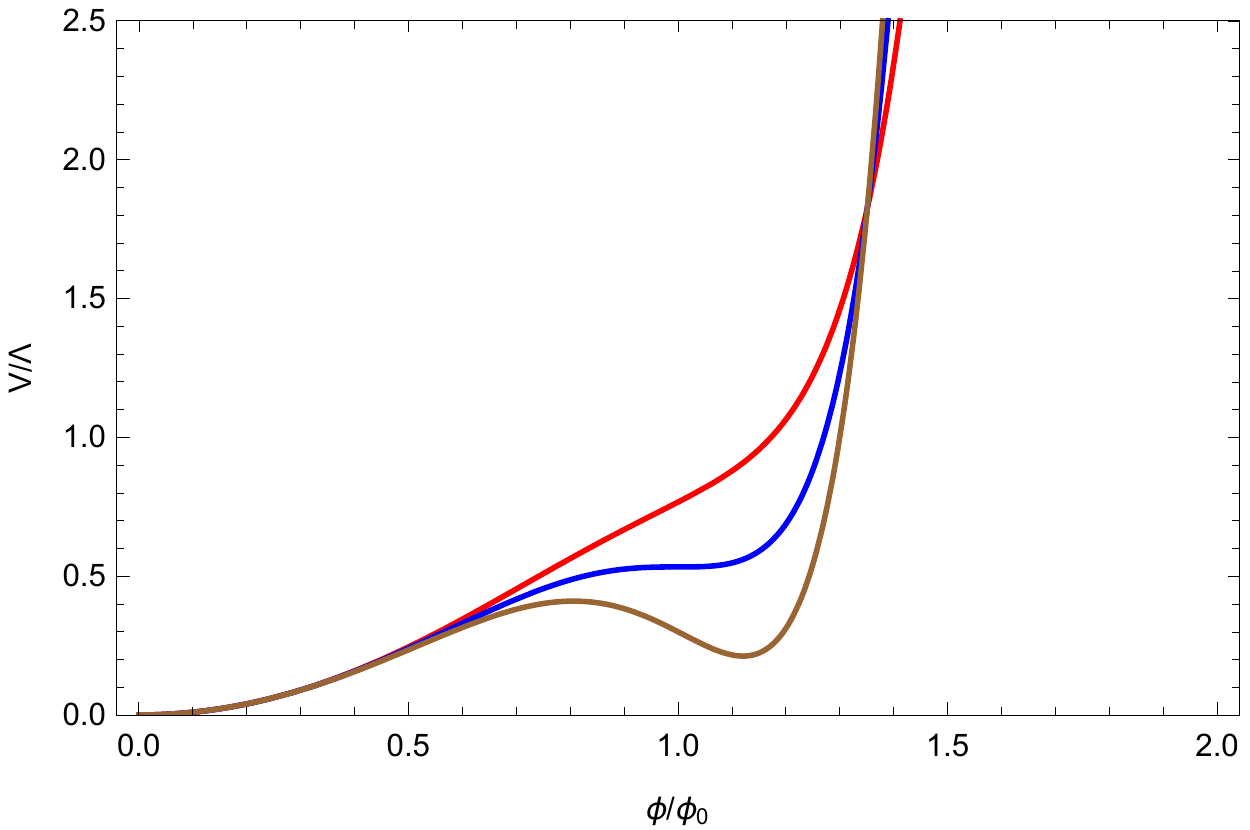}~~\includegraphics[scale=0.5]{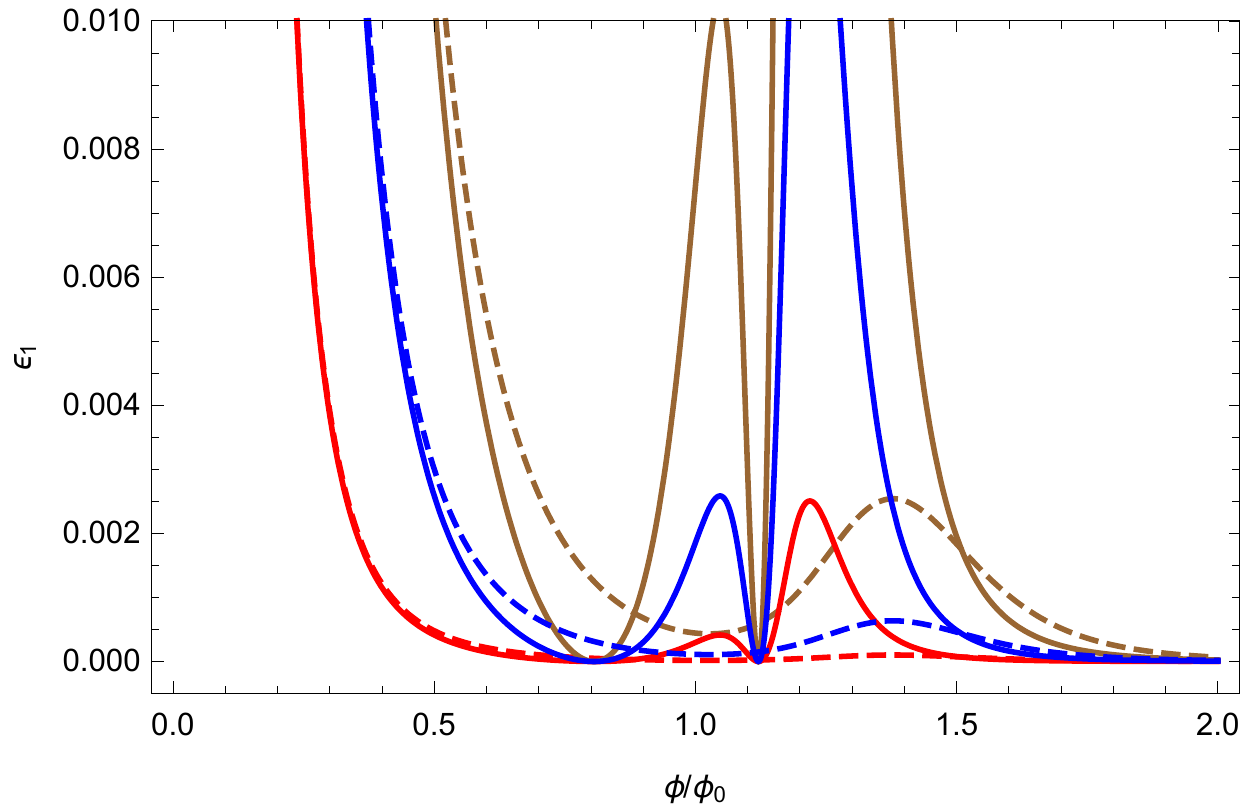}\\
  \includegraphics[scale=0.5]{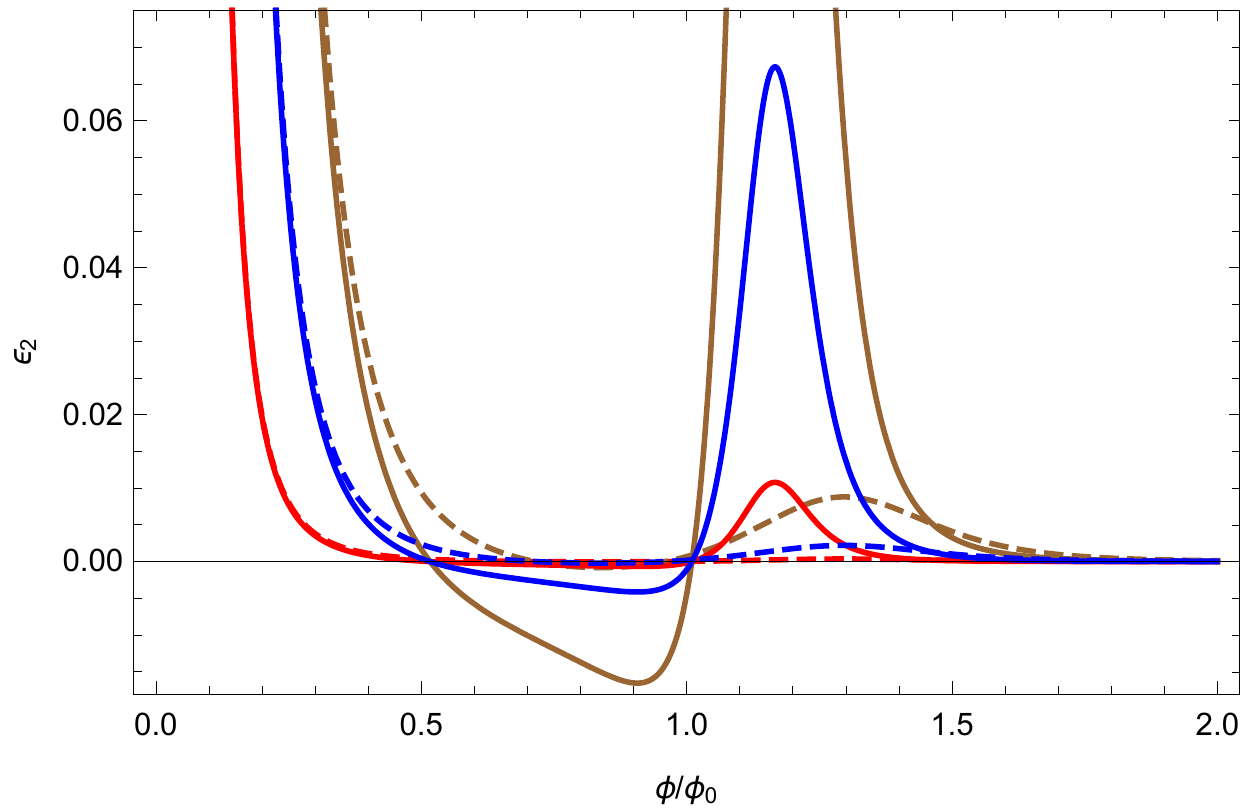}~~\includegraphics[scale=0.5]{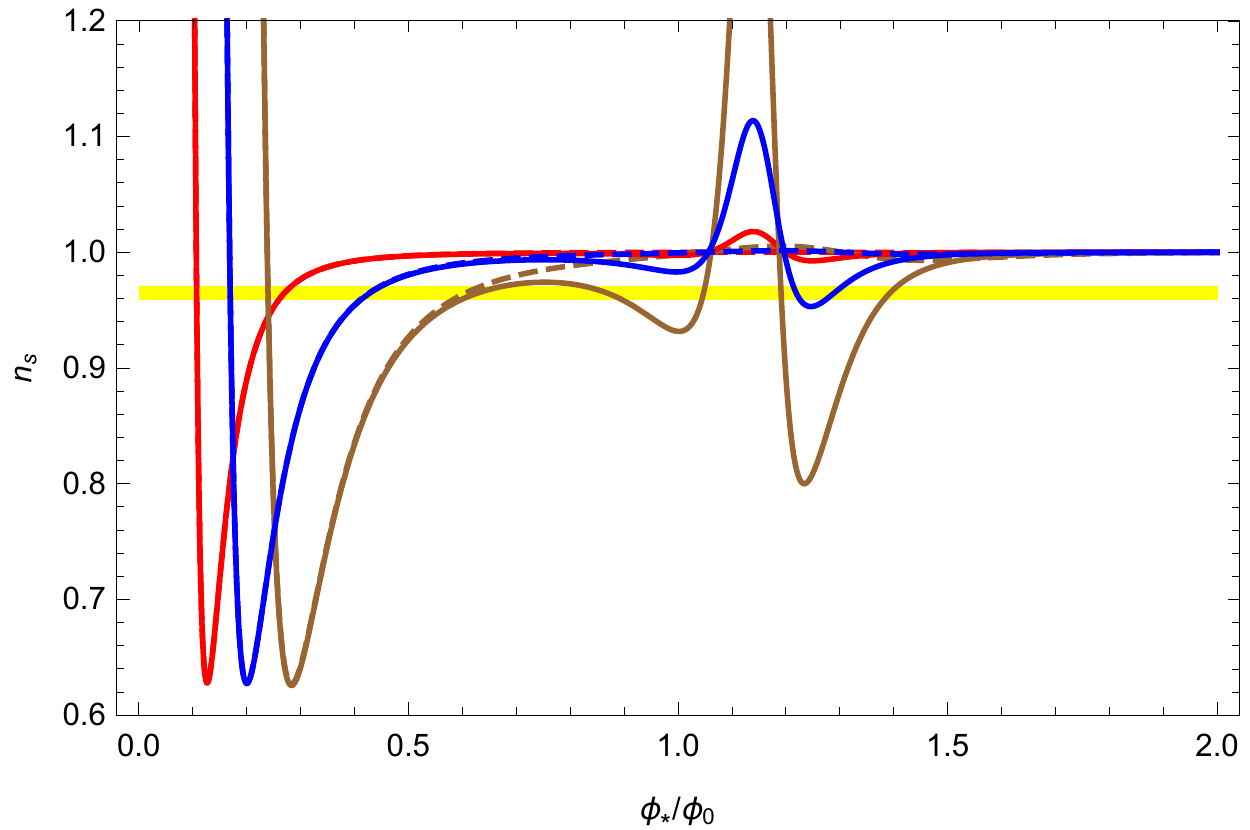}\\
  \caption{Field potential (top left) for the generalized MSSM inflation model for $\alpha = 0.5$ (red), $1$ (blue), and $1.5$ (brown). First and second slow-roll parameters (respectively top right and bottom left) and spectral index value as a function of $\phi_*$(bottom right), for the parameters $\phi_0 = 2.5 \Mpl$ (red), $\phi_0 = 1 \Mpl$ (blue); and $\phi_0 = 0.5 \Mpl$ (brown), and $\alpha = 1.5$ (solid) or $\alpha = 0.5$ (dashed). }\label{fig:pot_GMSSMI}
\end{figure}

\begin{figure}
  \centering
  \includegraphics[width=12cm]{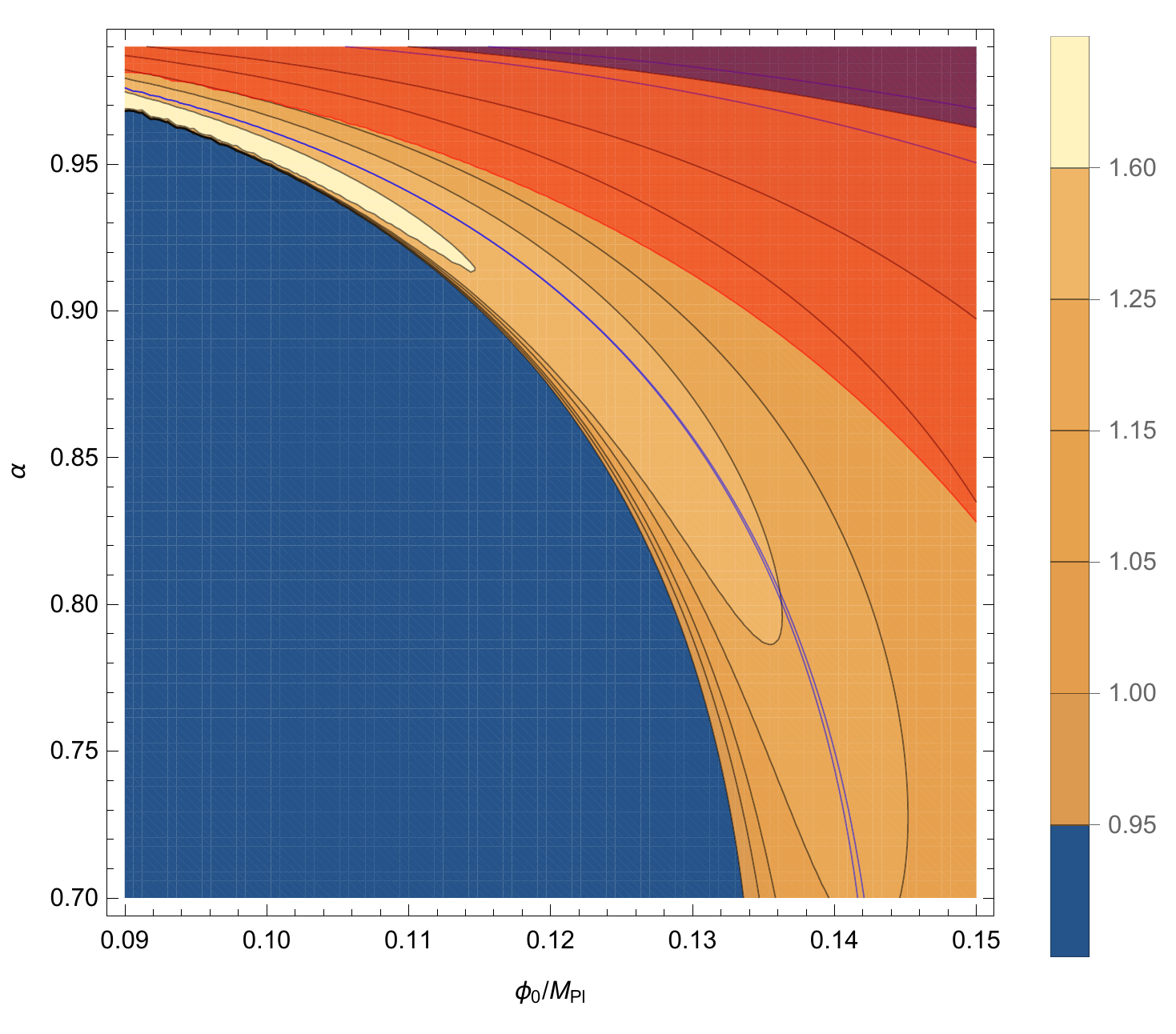}\\
  \caption{Spectral index values for the GMSSMI with $M = 0.01 \sqrt{\Lambda} / \Mpl$ at $k_d = 42 \Mpc^{-1}$ at $k_d = 42 \Mpc^{-1}$. The areas between two adjacent blue lines are consistent with the PLANCK constraints on scalar spectrum. The red region represents the allowed area for constraints on tensor modes. }\label{fig:ns_GMSSMI}
\end{figure}

The Minimal Supersymmetric Standard Model (MSSM) is a minimal extension of the Standard Model with copious cosmological consequences. In Ref. \cite{Allahverdi2006}, the authors argue that candidates for the inflaton can be provided by two combinations of flat directions, namely \textbf{LLe} and \textbf{udd}. The scalar potential for inflation can be parameterized as $V(\phi) = \Lambda \left[ (\frac{\phi}{\phi_0})^ 2 - \frac 2 3 (\frac{\phi}{\phi_0})^ 6 + \frac 1 5 (\frac{\phi}{\phi_0})^{10}   \right]\,$. As for the minimally coupled case, the MSSMI with the kinetic coupling also has an inflection point at $\phi = \phi_0$. Regions of red and blue spectrum are located on different sides of this point. However, in the classical slow-roll approximation, inflation approaches an infinite number of e-folds at the inflection point. The second criterion from Sec. \ref{sec:crit} thus cannot be applied. Therefore, we consider the Generalized MSSM Inflation (GMSSMI) scenario, where the inflection point is approximately. Following Ref. \cite{Martin2013a}, the potential of the GMSSMI model is parameterized as
\be
V(\phi) = \Lambda \left( x^ 2 - \frac 2 3 \alpha x^ 6 + \frac 1 5 \alpha x^{10}   \right)\,,
\ee
where $x \equiv \phi / \phi_0 $ and $\alpha$ is a dimensionless parameter that encodes the deviation from the MSSM inflation. If $\alpha > 1$, the potential develops a maximum. There are three possible inflationary field trajectories (towards both sides of the maximum, and from the large field regime towards decreasing field values). If $\alpha < 1$, the potential is monotonic. The field potential with different values of $\alpha$ is illustrated in Fig. \ref{fig:pot_GMSSMI}. The slow-roll parameters read
\begin{eqnarray}
\epsilon &=&\Mpl^4 \frac{M^2}{\Lambda} \frac{6750(\alpha x^4 (x^4 - 2)+ 1)^2}{\phi_0^2 x^4 \left(\alpha x^4 (3x^4 -10) +15 \right)^3}, \label{eq:eps1_GMSSMI}\\
\eta &=&\Mpl^4 \frac{M^2}{\Lambda} \frac{450(\alpha x^4 (9 x^4 - 10)+ 1)}{\phi_0^2 x^4 \left(\alpha x^4 (3x^4 -10) +15 \right)^2}, \label{eq:eps2_GMSSMI}
\end{eqnarray}
where the slow roll approximation $3H^2 = V$ is applied. In the MSSM scenario, it suggests that $\Lambda\sim\Mpl^{3/2}m_{\rm SUSY}^{5/2}$ and $\phi_0\sim\Mpl^{3/4}m_{\rm SUSY}^{1/4}$. For the minimally coupled case, inflation can only proceed in a fine-tuned region of $\alpha \simeq 1$, to satisfy the minimum of the first slow-roll parameter $\epsilon_{min} < 1$, and to create enough e-folds (c.f. Ref. \cite{Martin2013a}). With the non-minimal derivative coupling, the slow-roll parameters are suppressed, whereas the e-folds number is enhanced by a factor of $3H^2/M^2$. In the high friction limit where $M^2 \ll H^2$, the allowed region of $\alpha$ is much larger. We can still set the coupling constant $M = 0.01 \sqrt{\Lambda} / \Mpl$. The slow-roll parameters and the spectral index are then presented in Fig. \ref{fig:pot_GMSSMI}. Inflation ends by violation of the slow-roll conditions when $\eta > 1$. The corresponding end point is $x_{end} = \sqrt[4]{0.0002 \Mpl^2 / \phi_0^2}$. For applying the third criterion from Section \ref{sec:crit}, we illustrate the spectral index in Fig. \ref{fig:ns_GMSSMI} in the plane ($\phi_0$,$\alpha$) for the two pivot scales $k_p = 0.05 \Mpc^{-1}$ and $k_d = 42 \Mpc^{-1}$. The potentially interesting region is the area between two adjacent blue lines and $\alpha < 0.97$ in the figure. After evaluating the value of the tensor-to-scalar ratio at $k = 0.002 \Mpc^{-1}$, we find that parameter sets in the potentially interesting region cannot satisfy the constraints on tensor modes.

\begin{figure}
  \centering
  \includegraphics[width=12cm]{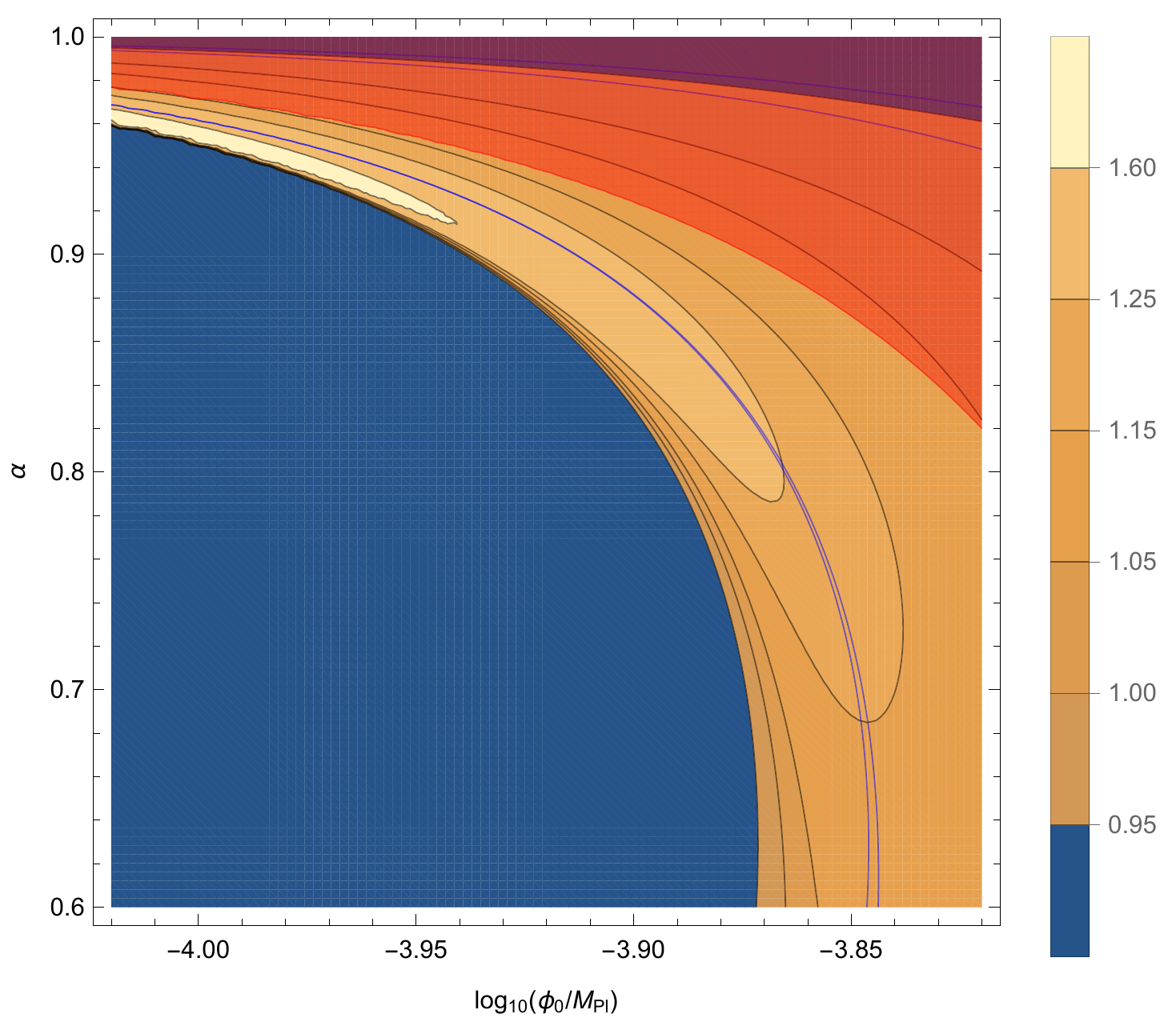}\\
  \caption{Same as in Fig. \ref{fig:ns_GMSSMI} with $M = 10^{-5} \sqrt{\Lambda} / \Mpl$.}\label{fig:ns2_GMSSMI}
\end{figure}

Nevertheless, compared with the minimally coupled case in Ref. \cite{Clesse:2014pna}, the value of parameter $\phi_0$ in the potentially interesting region and the corresponding values of inflaton field can be suppressed to be sub-Planckian. In Ref. \cite{Allahverdi2006}, it suggests $\phi_0 \simeq 10^{-4} \Mpl$. In order to be in agreement with the MSSM scenario, one needs to turn on an extremely strong kinetic coupling, such as $M = 10^{-5} \sqrt{\Lambda} / \Mpl$. In this case, the end point of inflation is $x_{end} = 3.76 \times 10^{-3} \sqrt{\Mpl / \phi_0}$. The corresponding plane diagram of the value of the spectral index is presented in Fig \ref{fig:ns2_GMSSMI}. The area between two adjacent blue lines and $\alpha < 0.97$ is the potentially interesting parameter regime in the scope of primordial scalar fluctuations.
If one choose a point inside the region $n_s(k_d) > 1.25$, such as $\log_{10} (\phi_0 / \Mpl) = -3.87$ and $\alpha = 0.818$, the resulting CMB distortion will be $\mu = 5.07 \times 10^{-8}$ (evaluated by a modified version of the Idistort code \cite{Khatri2012a}) which can be detected at $5 \sigma$. Moreover, the field value at horizon exit at $60$ e-folds reads $\phi (N_k = 60) \approx 1.25 \phi_0 = 1.69 \times 10^{-4} \Mpl$. Hence the inflaton field runs on sub-Planckian scales during the inflation process. However, the potentially interesting regime is still not allowed by the constraints on tensor modes, as it do not overlap with the red region in Fig \ref{fig:ns2_GMSSMI}. Compared with the $M = 0.01 \sqrt{\Lambda} / \Mpl$ case above, it is easy to find that the mass-dimension model parameter $\phi_0$ varies proportionally when the coupling $M$ changes. According to \eqref{eq:index}, \eqref{eq:ratio}, \eqref{eq:eps1_GMSSMI} and \eqref{eq:eps2_GMSSMI}, the behaviors of the scalar spectral index and the tenor-to-scalar ration are therefore similar for different values of $M$. We can thus conclude that the GMSSMI model does not lead to any observable distortion of the CMB frequency spectrum.

\section{Generalized Renormalizable Inflection Point Inflation (GRIPI)}
\begin{figure}
  \centering
\includegraphics[scale=0.5]{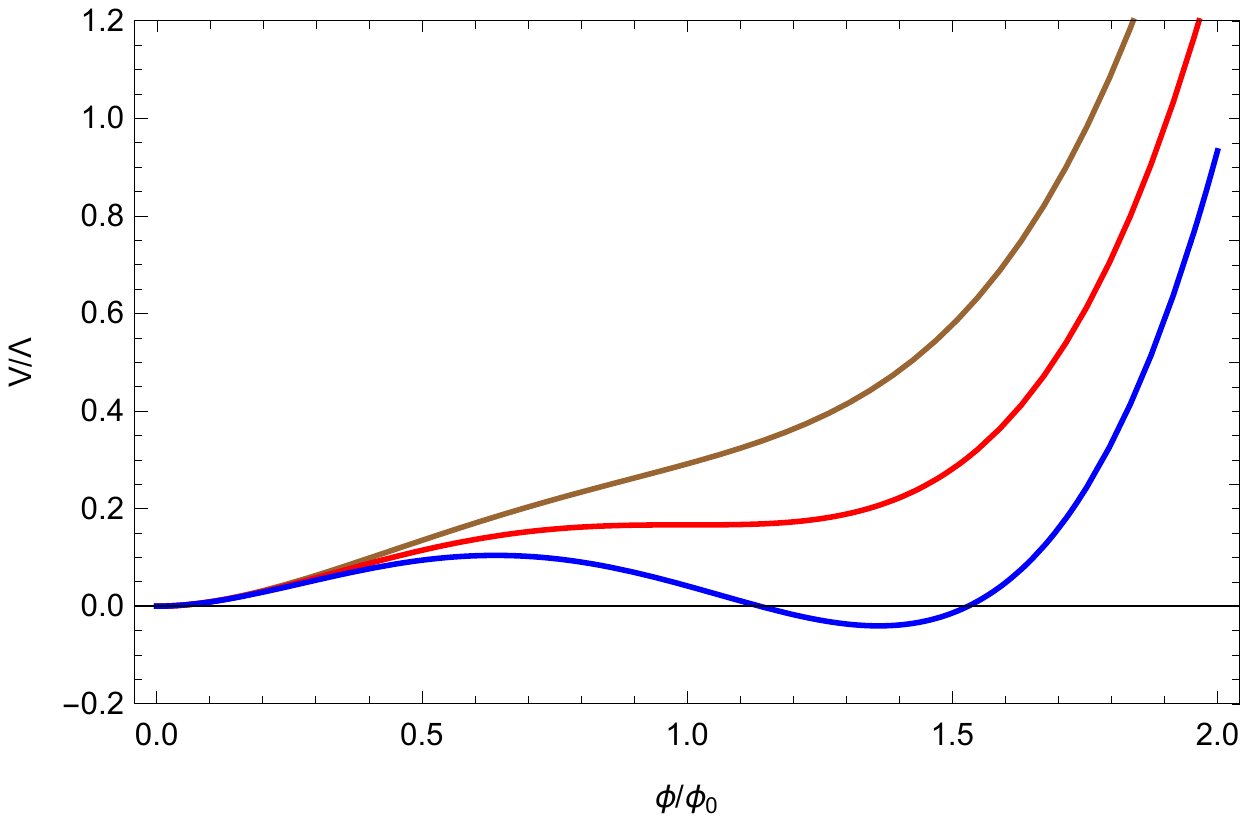}~~\includegraphics[scale=0.5]{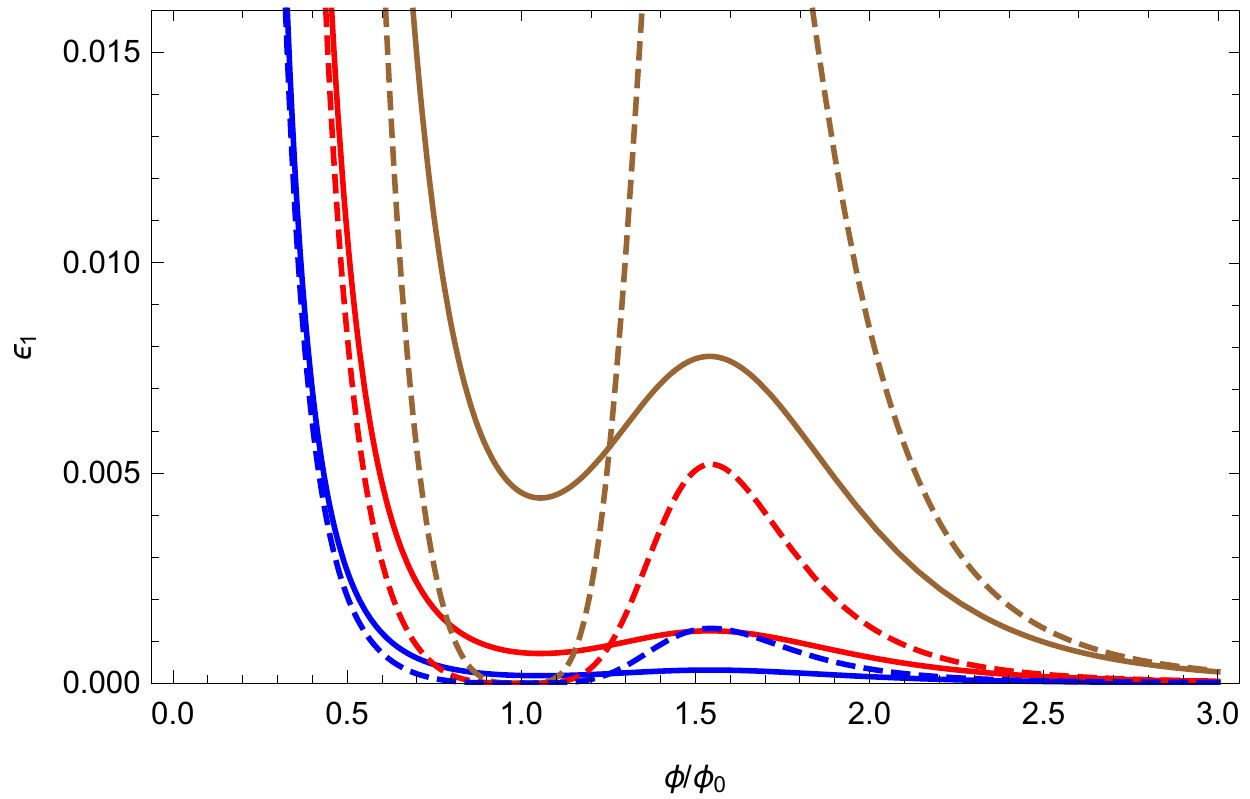}
\includegraphics[scale=0.5]{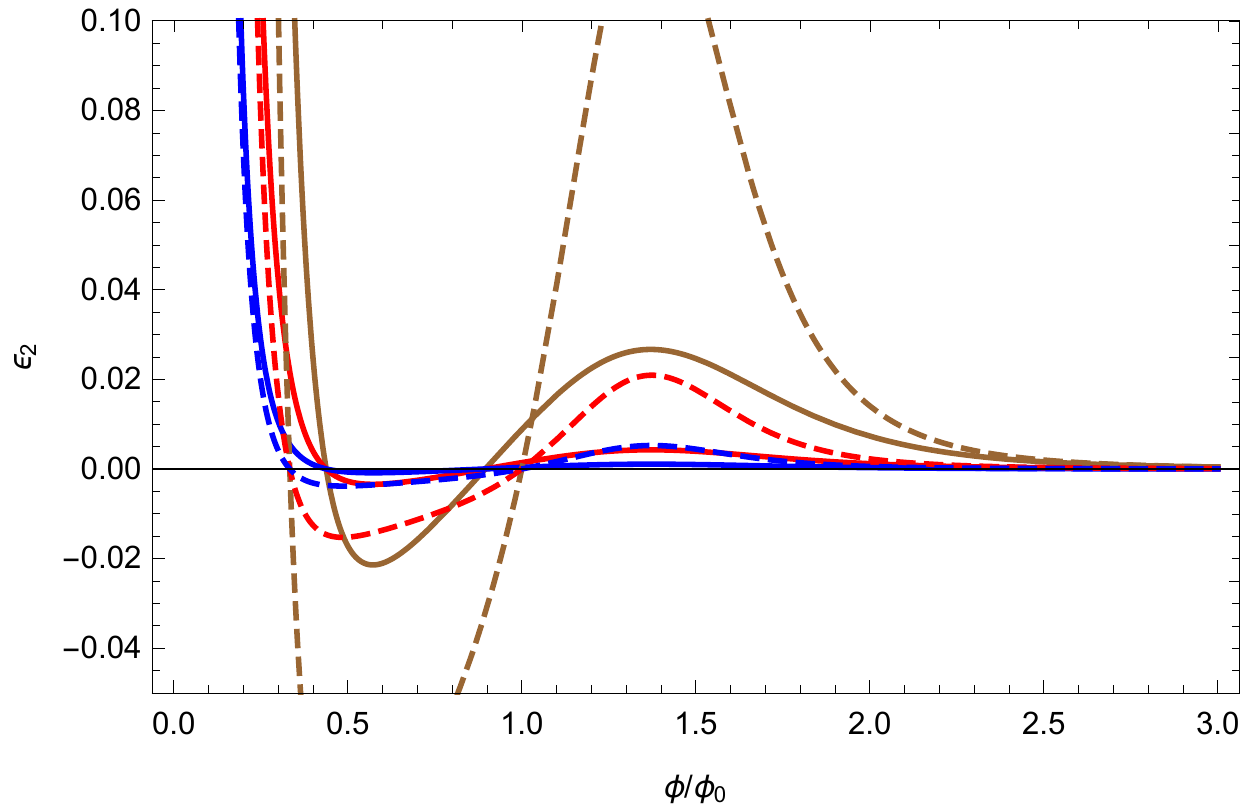}~~\includegraphics[scale=0.5]{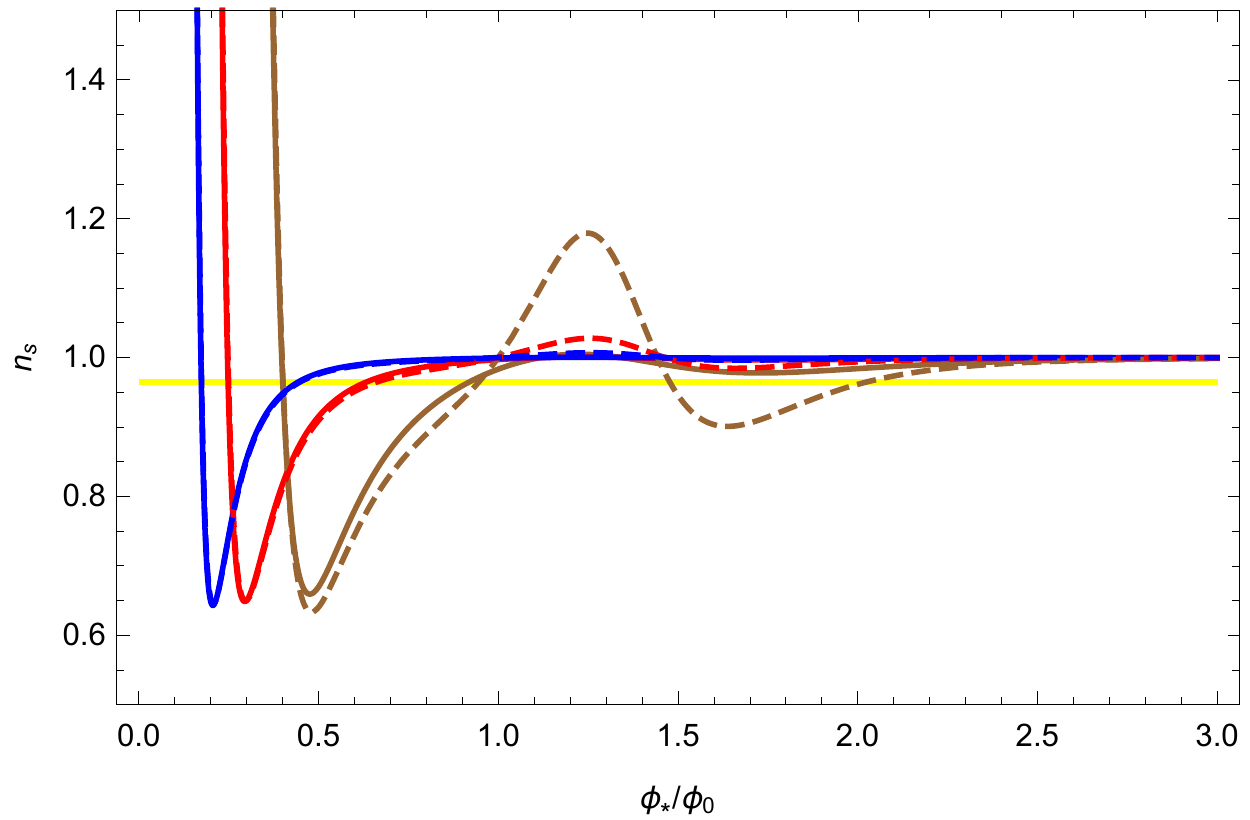}
  \caption{Generalized Renormalizable Inflection Point Inflation (GRIPI) potential (top left) for $\alpha = 0.85$ (brown), $\alpha = 1$ (red), and $\alpha = 1.15$ (blue). First and second slow-roll parameters (respectively top right and bottom left) and spectral index value as a function of $\phi_*$ (bottom right), for $\phi_0 = (0.2,0.5,1) \Mpl$ (respectively brown, red and blue) and $\alpha = 0.85 ,1 $ (respectively solid and dashed).}\label{fig:pot_GRIPI}
\end{figure}

\begin{figure}
  \centering
  \includegraphics[width=12cm]{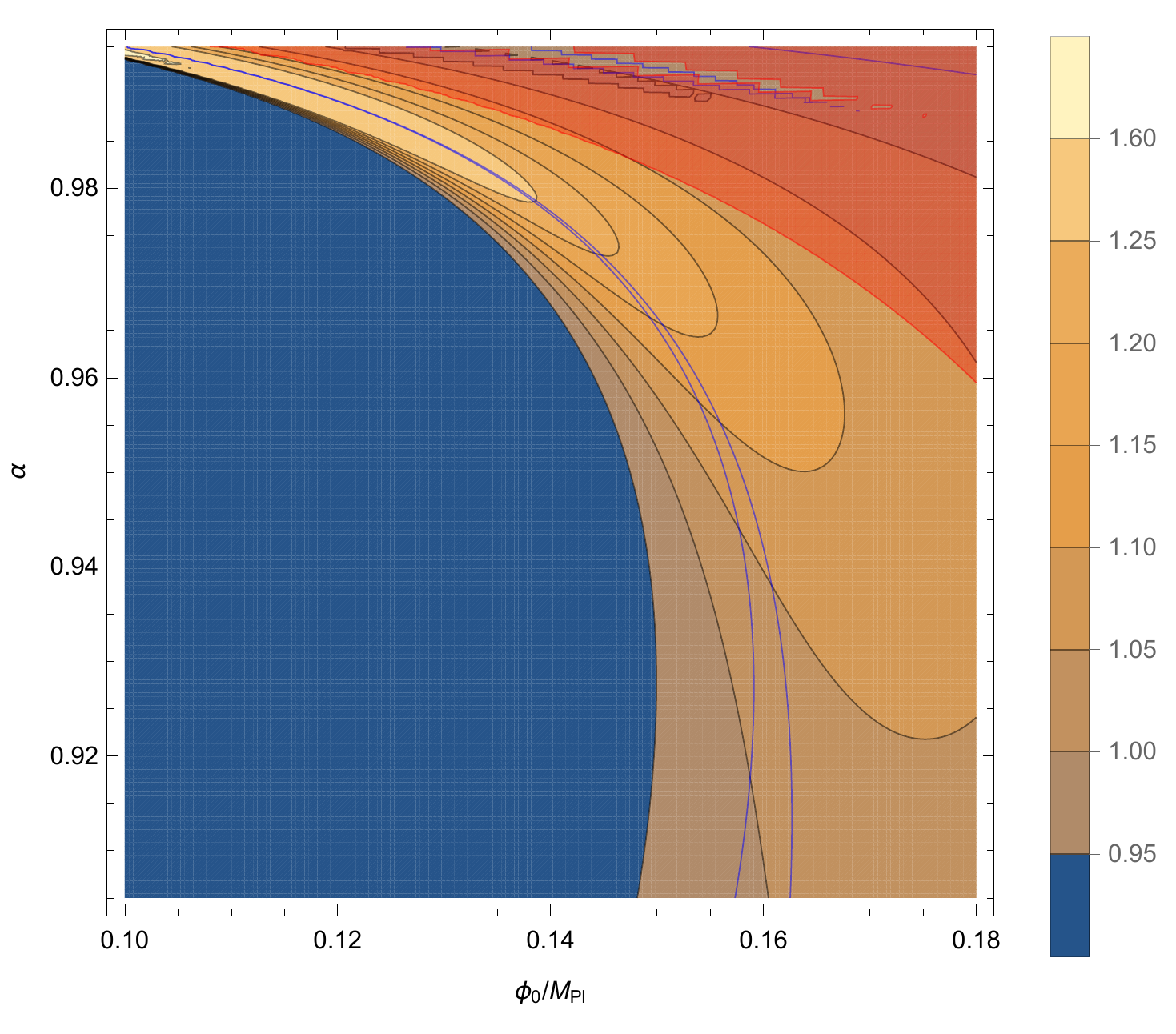}
  \caption{Spectral index value in the parameter space ($\phi_0$, $\alpha$) for GRIPI with the kinetic coupling $M = 0.01 \sqrt{\Lambda} / \Mpl$, for the pivot scales of CMB distortions $k_d = 42 \Mpc^{-1}$. The areas between the adjacent solid blue lines are consistent with PLANCK constraints on scalar spectrum. The red region is consistent with constraints on tensor modes.} \label{fig:ns_GRIPI}
\end{figure}

\begin{figure}
  \centering
  \includegraphics[width=12cm]{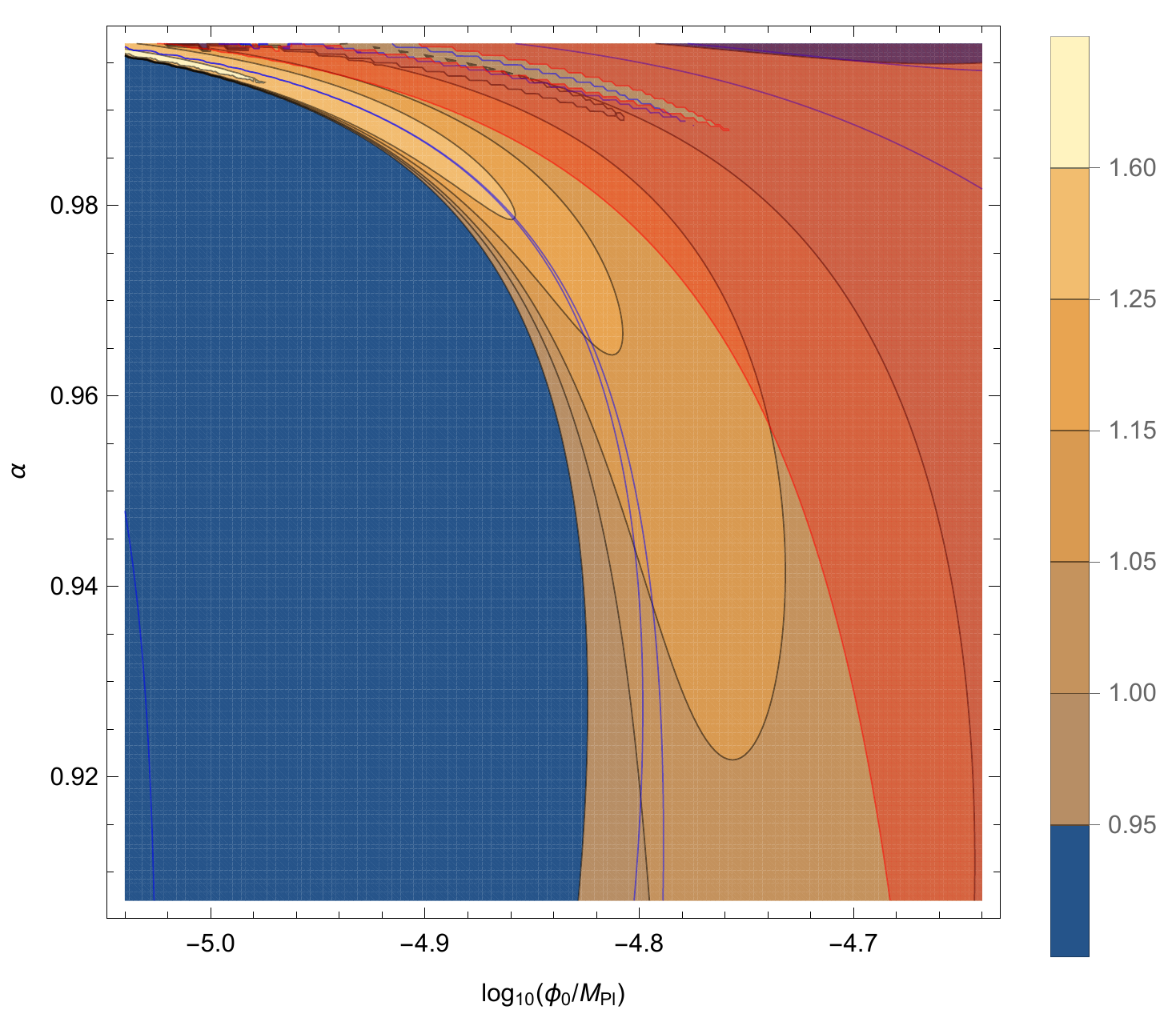}
  \caption{Same as in Fig. \ref{fig:ns_GRIPI} with the coupling constant $M = 10^{-6} \sqrt{\Lambda} / \Mpl$.}\label{fig:ns2_GRIPI}
\end{figure}

As for MSSM inflation, the renormalizable infection point inflation features an exact inflection point to cause an eternal inflation. The number of e-folds diverges at this point. It immediately leads us to Generalized Renormalizable Inflection Point Inflation (GRIPI), where the inflection point is approximate. The GRIPI is similar to GMSSMI, and differs only at the powers of the inflaton. The field potential can be expressed as
\be
V(\phi)=\Lambda\left(x^2-\frac{4}{3}\alpha{x^3}+\frac{1}{2}\alpha{x^4}\right),
\ee
where $x = \phi / \phi_0$, and $\alpha$ is a dimensionless parameter. The field potential is presented in Fig. \ref{fig:pot_GRIPI}. When $\alpha < 1$, it is a monotonically increasing function of the field. As for GMSSMI, when $\alpha > 1$, the potential develops a maximum with three possible inflationary regimes.
The slow-roll parameters are given by
\begin{eqnarray}
  \epsilon &=& \Mpl^4 \frac{M^2}{\Lambda}\frac{(2x-4x^2\alpha+2x^3\alpha)^2}{2{\phi_0}^2x(x^2 - \frac{4x^3\alpha}{3} + \frac{x^4\alpha}{2})^3 }, \\
  \eta &=& \Mpl^4 \frac{M^2}{\Lambda} \frac{72(1+x(-4+3x)\alpha)}{{\phi_0}^2x^4(6+x(-8+3x)\alpha)^2}.
\end{eqnarray}
Since the rescaled field $x$ is dimensionless, we can fix the coupling constant as $M = 0.01 \sqrt{\Lambda} / \Mpl$, such that inflation ends by violation of slow-roll conditions at $x_{end} = \sqrt[4]{2 \times 10^{-4} \Mpl^2 / \phi_0^2}$. Then we can obtain the field value $\phi_k$ for each given mode at horizon exit by numerically integrating the Klein-Gordon equation in the slow-roll approximation. The spectral index predictions in the two-dimensional parameter space for the pivot scale $k_d = 42 \Mpc^{-1}$ are illustrated in Fig. \ref{fig:ns_GRIPI}. The band between two blue lines in the center of the figure is the available region consistent with PLANCK scalar spectrum. The potentially interesting region is the overlap between this band and the region leading to $n_s >1$ on CMB distortion scales. Imposing the $95\%$ CL constraints on tensor modes, we find that there is no overlap with the potentially interesting region. Then, we look into the case of extremely strong coupling by setting the coupling constant to be $M = 10^{-6} \sqrt{\Lambda} / \Mpl$. The corresponding plane plot for the value of spectral index $n_s$ at $k_d = 42 \Mpc^{-1}$ is shown in Fig. \ref{fig:ns2_GRIPI}. One may note that the value of parameter $\phi_0$ is strongly suppressed to be $\phi_0 \simeq 10^{-5} \Mpl$ which is in agreement with the MSSM scenario \cite{Hotchkiss2011}. However, there is still no overlap between parametric regions in agreement with the PLANCK observation and regions associated with an increase of power at the scale $k_d$. We can therefore conclude that the model cannot lead to any observable spectral distortion of the CMB spectrum.

\section{Running-mass Inflation (RMI)}
\label{sec:RMI}
\begin{figure}
  \centering
  \includegraphics[scale=0.6]{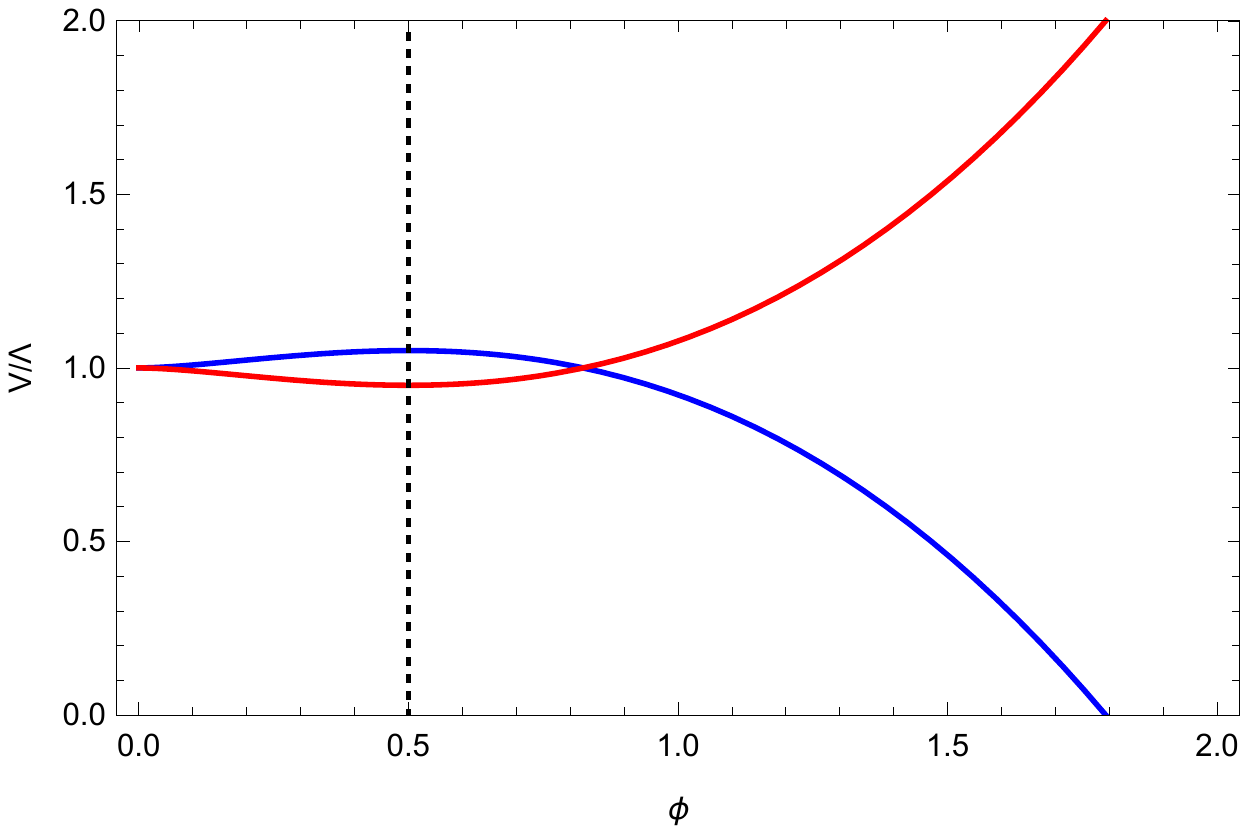}~~\includegraphics[scale=0.6]{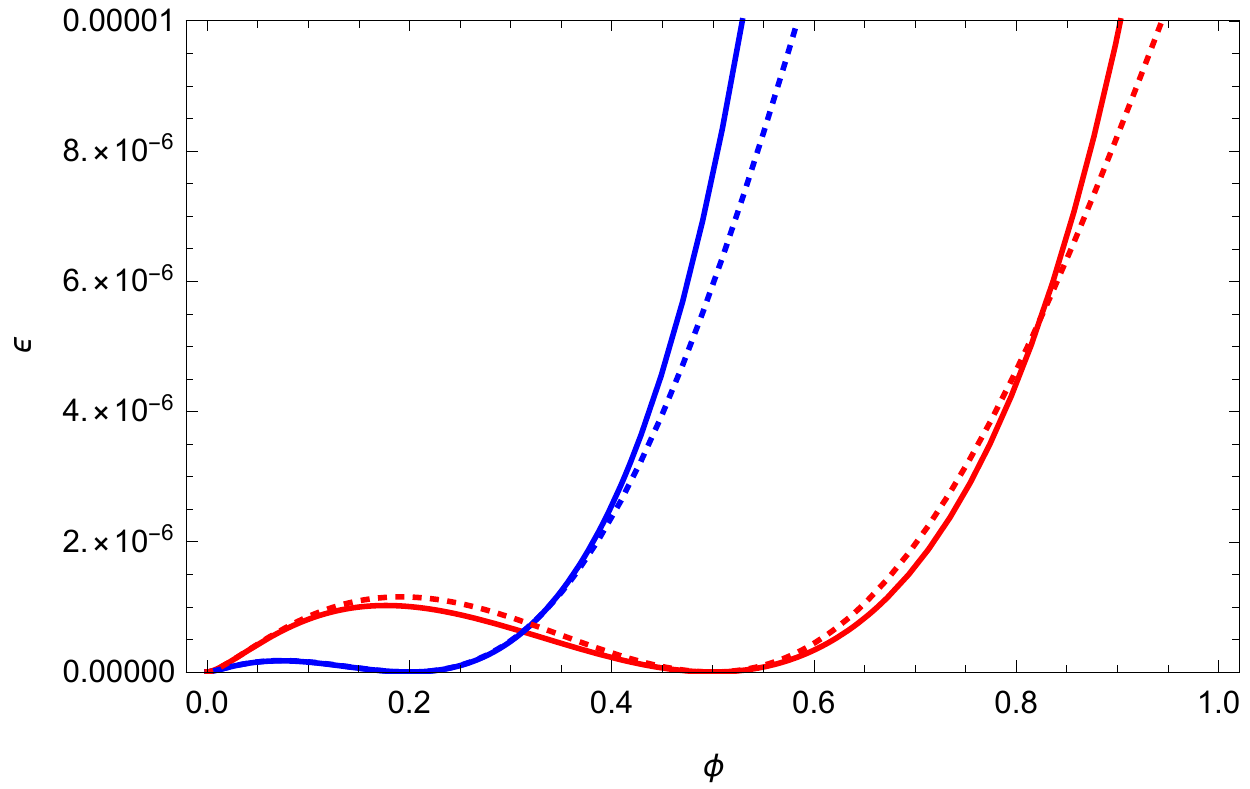}
  \includegraphics[scale=0.6]{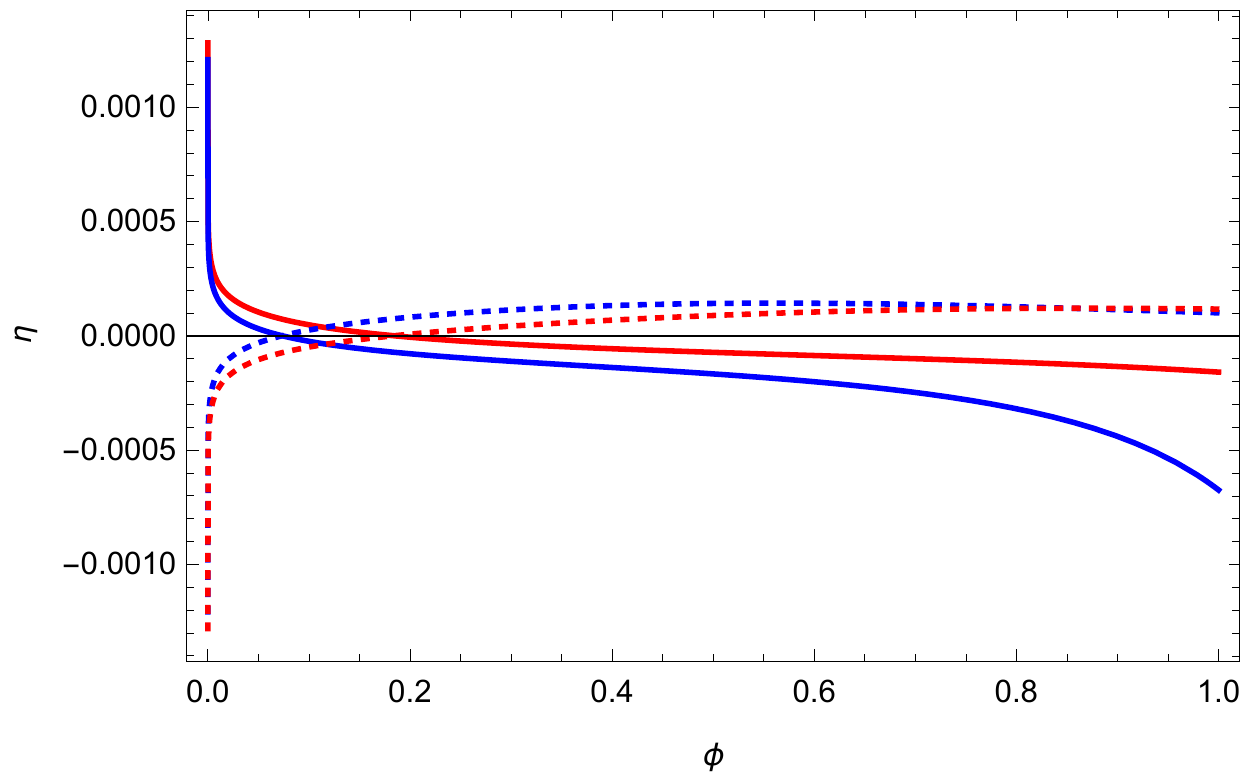}~~\includegraphics[scale=0.6]{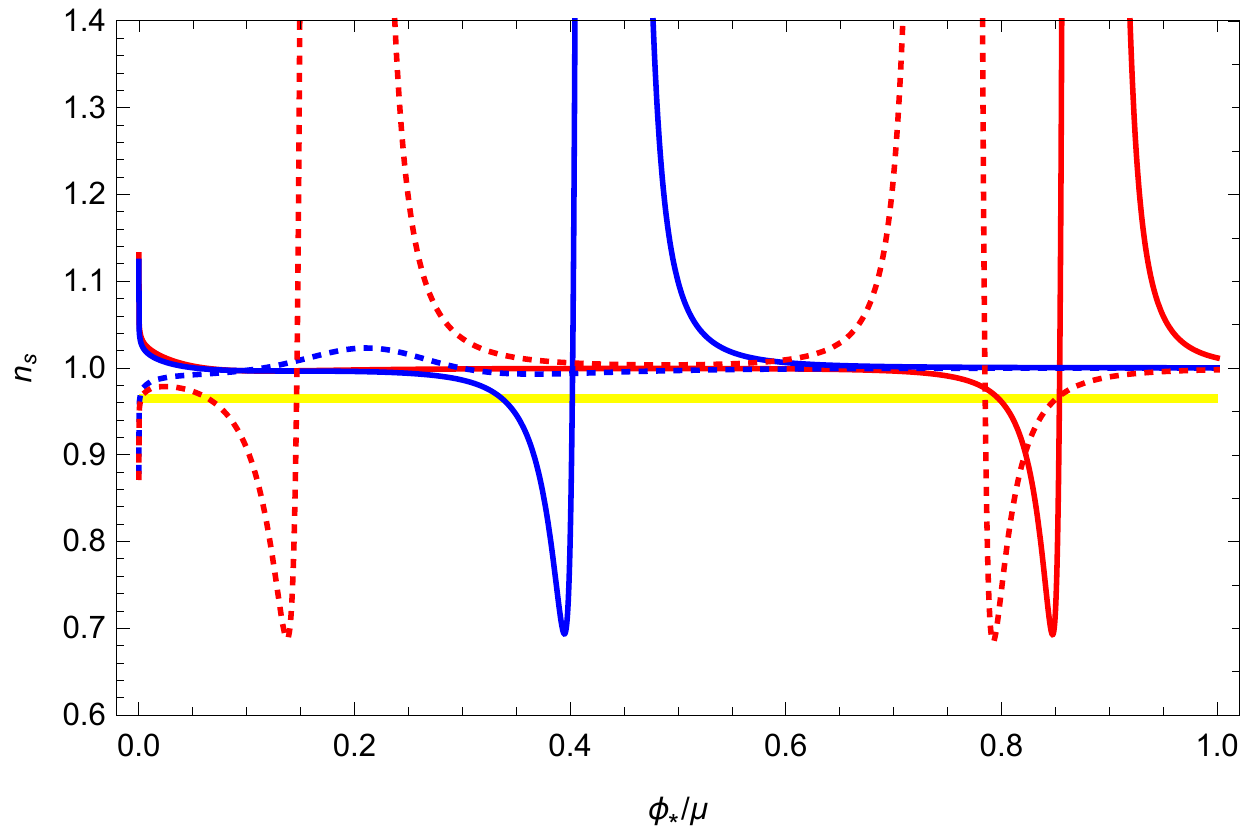}
  \caption{Top left: field potential for the running-mass Inflation (RMI) for $\phi_0 = 0.5 \Mpl$, $c = 0.8 $ (blue) and $c = -0.8$ (red). Top right and bottom left, respectively: first and second slow-roll parameter for the parameters $\phi_0 = 0.5 \Mpl$ (red), $\phi_0 = 0.2 \Mpl$ (blue), and $c = 0.8 $ (solid) and $c = -0.8$ (dotted). Bottom left: spectral index value as a function of $\phi_*$ for the parameters $\phi_0 = 0.5 \Mpl$ (red), $\phi_0 = 0.2 \Mpl$ (blue); and $c = 40 $ (solid), $c = -40$ (dotted). }\label{fig:pot_RMI}
\end{figure}

Running-mass Inflation (RMI) also comes from a supersymmetric framework \cite{Martin2013a,Stewart1997,Stewart1997a,Covi1999,Covi1999a}. A flat direction of the potential is lifted when the supersymmetry is explicitly broken by a soft term. It gives a logarithmic correction to the tree-level inflaton mass. Then the potential reads
\be
V(\phi)=\Lambda \left[1-\frac{c}{2} \left(-\frac{1}{2} + \ln{\frac{\phi}{\phi_0}}\right) \frac{\phi^2}{M_{pl}^2} \right],
\ee
where $\phi_0$ is the renormalization scale and $c$ is a dimensionless parameter. The value of $c$ can be either positive or negative. These different possibilities are illustrated in Fig. \ref{fig:pot_RMI}. If $c > 0$, the potential develops a maximum at $\phi = \phi_0$. There are two possible inflationary regimes: (i) from the maximum towards the decreasing field values (RMI1); (ii) from the maximum towards the increasing field values (RMI2). If $c < 0$, there is a minimum located at $\phi = \phi_0$ with two inflationary trajectories. Inflation can proceed from the small field values regime towards the minimal (RMI3) or from the large field values towards the minimum (RMI4). The slow-roll parameters are given by
\begin{eqnarray}
  \epsilon &=& \Mpl^2 \frac{M^2}{\Lambda} \frac{32 c^2 x^2 \log^2[\frac{x}{\phi_0}]}{ (4+ c x^2 - 2 c x^2 \log[\frac{x}{\phi_0}])^3} \\
  \eta &=& - \Mpl^2 \frac{16cM^2(1+\log[\frac{x}{\phi_0}] )}{\Lambda (4 + c x^2-2c x^2 \log[\frac{x}{\phi_0}] )^2 }
\end{eqnarray}
where $x \equiv \phi / \Mpl$. We set the derivative coupling $M = 0.01 \sqrt{\Lambda} / \Mpl$. The slow-roll parameters and spectral index value as a function of $\phi_*$ are demonstrated in Fig. \ref{fig:pot_RMI}. As for the original model \cite{Clesse:2014pna}, the RMI2 is irrelevant in the scope of the present paper, since the spectral index is always red. The other three regimes (RMI1, RMI3, and RMI4) allow one to satisfy observation constraints of $n_s$ from the CMB anisotropies, and give rise to an enhanced spectrum amplitude on smaller scales. We then study these regimes in detail by integrating the slow-roll dynamics numerically.

\subsection{RMI1: \texorpdfstring{$c>0$}{c>0}, \texorpdfstring{$\phi < \phi_0$}{phi<phi0}}
\begin{figure}
  \centering
  \includegraphics[width=12cm]{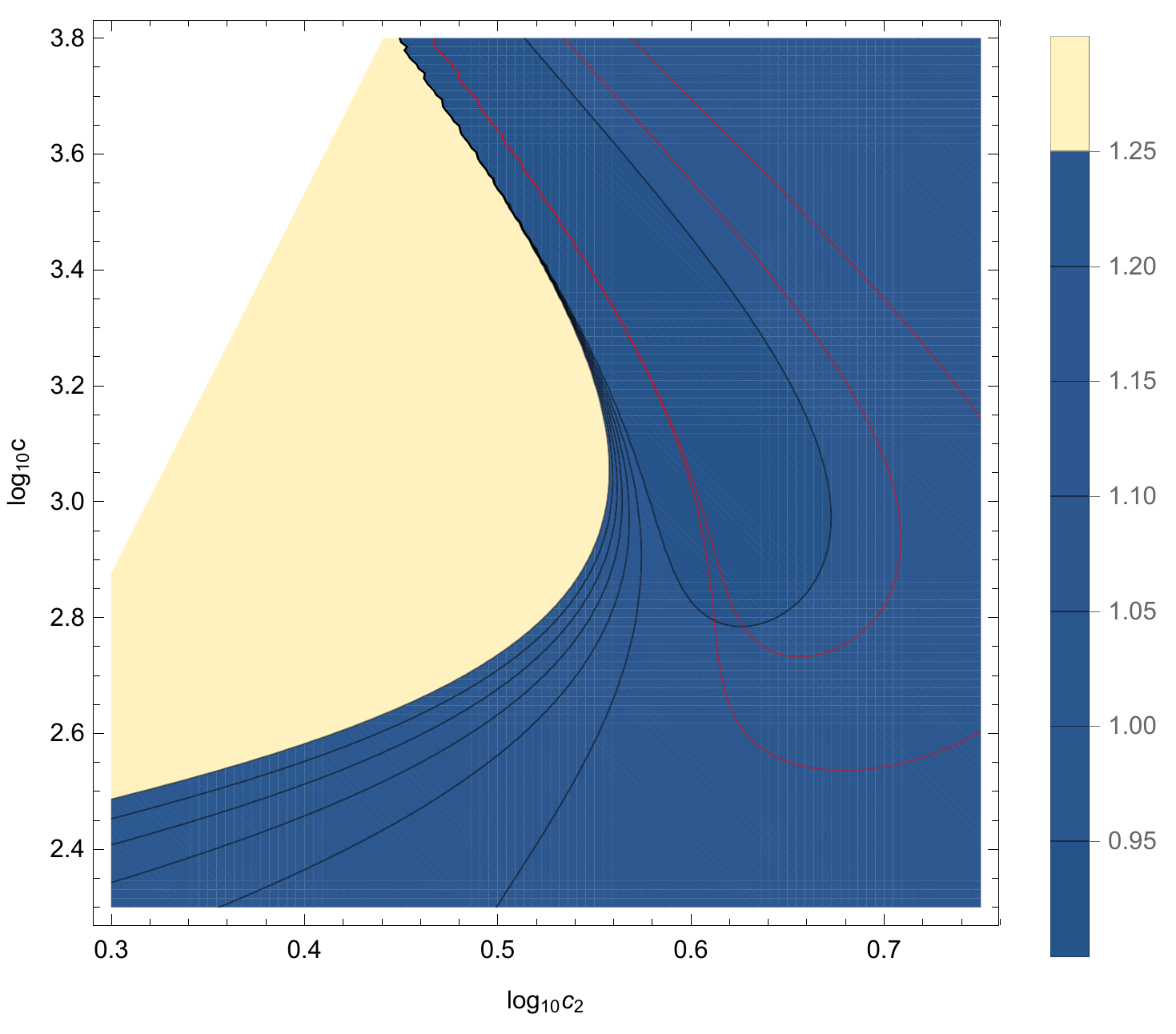}
  \caption{Spectral index value for the RMI1 regime with $\phi_0 = 0.1 \Mpl$ at $k_d = 42 \Mpc^{-1}$. The area between the red lines is allowed by the PLANCK constraints at the pivot scale of CMB anisotropies $k_p = 0.05 \Mpc^{-1}$. }\label{fig:ns_RMI1}
\end{figure}

As mentioned in Ref. \cite{Martin2013a}, for the validity of the RMI model, inflation must end by a tachyonic waterfall instability. In the RMI1 regime, we parameterize the critical instability point as $\phic = c_2 \phi_0$, where $c_2 < 1$ is a dimensionless parameter. In order to obtain the scalar power spectrum, we numerically integrate the Klein-Gordon equation in the slow-roll approximation. Spectral index values at the pivot scale of CMB distortions, $k_d = 42 \Mpc^{-1}$ in the parameter space ($c_2$,$c$), are shown in Fig. \ref{fig:ns_RMI1} for the case of $\phi_0 = 0.1 \Mpl$. The area enclosed by red solid lines is in agreement with PLANCK observations. This area does not overlap with regions where $n_s > 1$ at the scale $k_d$. Therefore, the RMI1 regime cannot satisfy the PLANCK constraints and gives rise to a detectable spectral distortion simultaneously.

\subsection{RMI3 and RMI4: \texorpdfstring{$c < 0$}{c<0}, \texorpdfstring{$\phi < \phi_0$}{phi<phi0} or \texorpdfstring{$\phi > \phi_0$}{phi>phi0}}
\begin{figure}
  \centering
  \includegraphics[width=9cm]{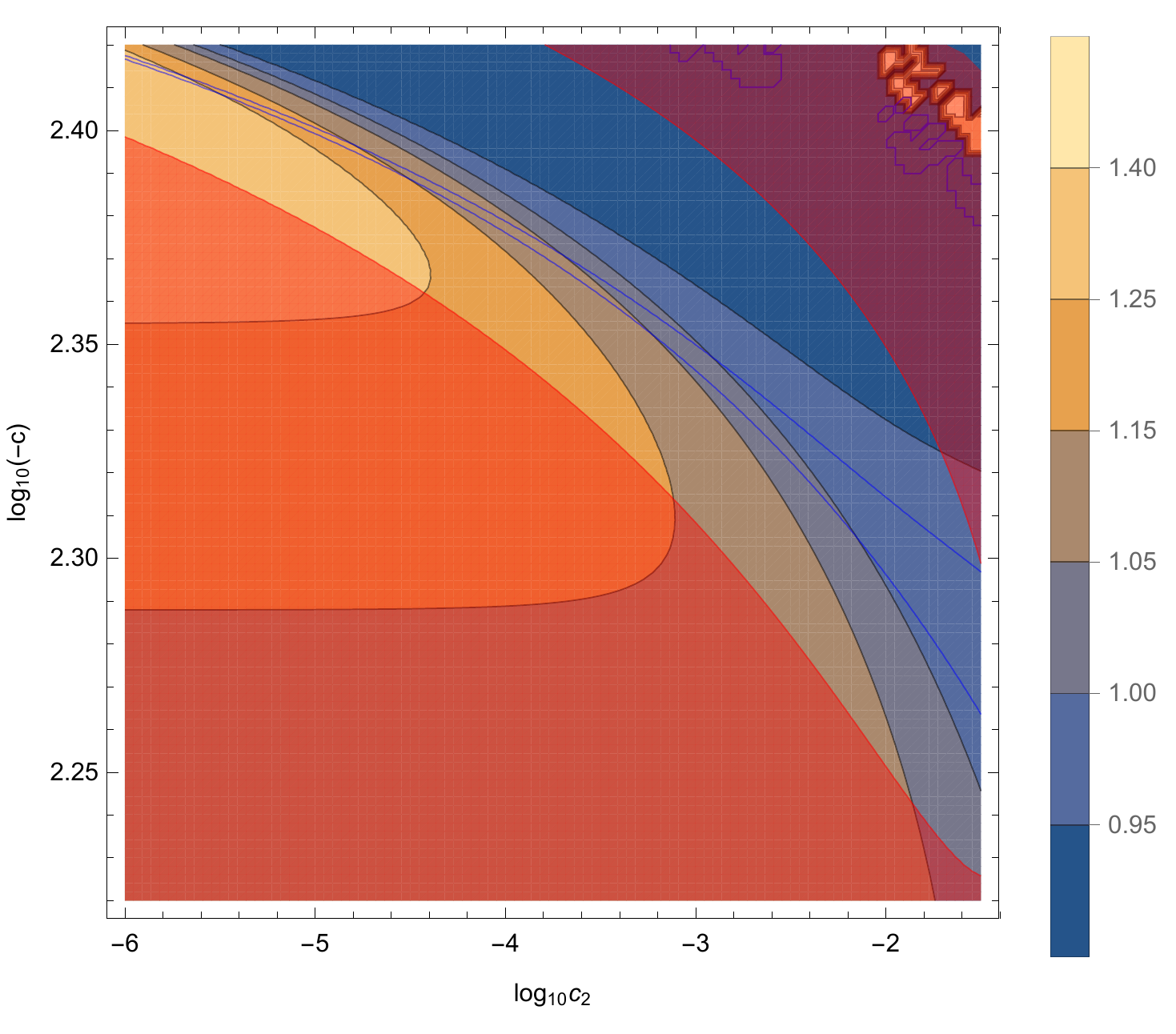}~~\includegraphics[width=9cm]{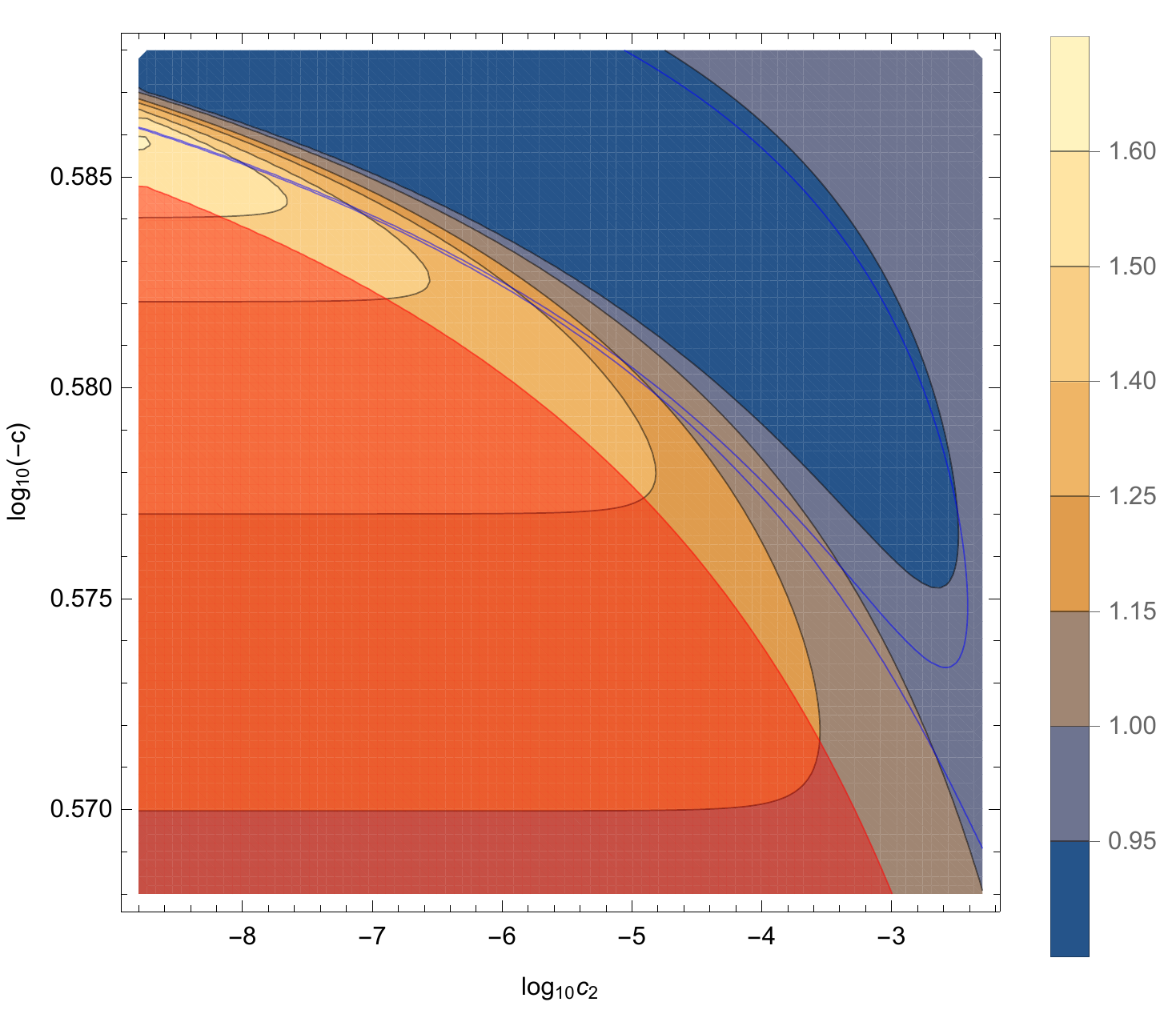}
  \caption{Spectral index value for the RMI3 regime with $\phi_0 = 0.1 \Mpl$ (left) and $\phi_0 = 1 \Mpl$ (right) at $k_d = 42 \Mpc^{-1}$. The area between the blue lines is allowed by the PLANCK constraints on scalar spectrum at the pivot scale of CMB anisotropies $k_p = 0.05 \Mpc^{-1}$. The red region is consistent with the constraints on primordial gravitational waves. The noises caused by numerical instability on the top right can be ignored.}\label{fig:ns_RMI3}
\end{figure}

\begin{figure}
  \centering
  \includegraphics[width=9cm]{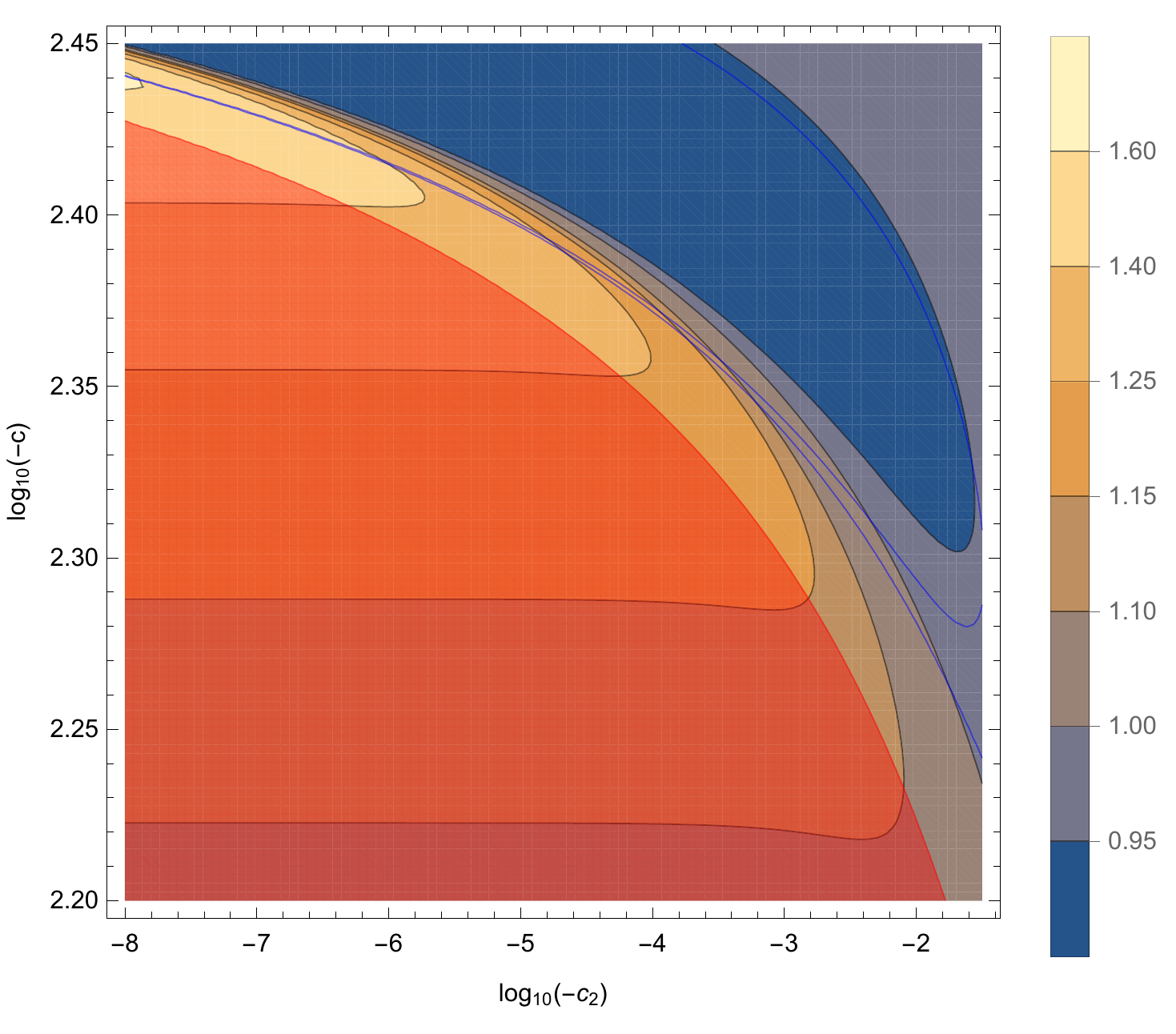}~~\includegraphics[width=9cm]{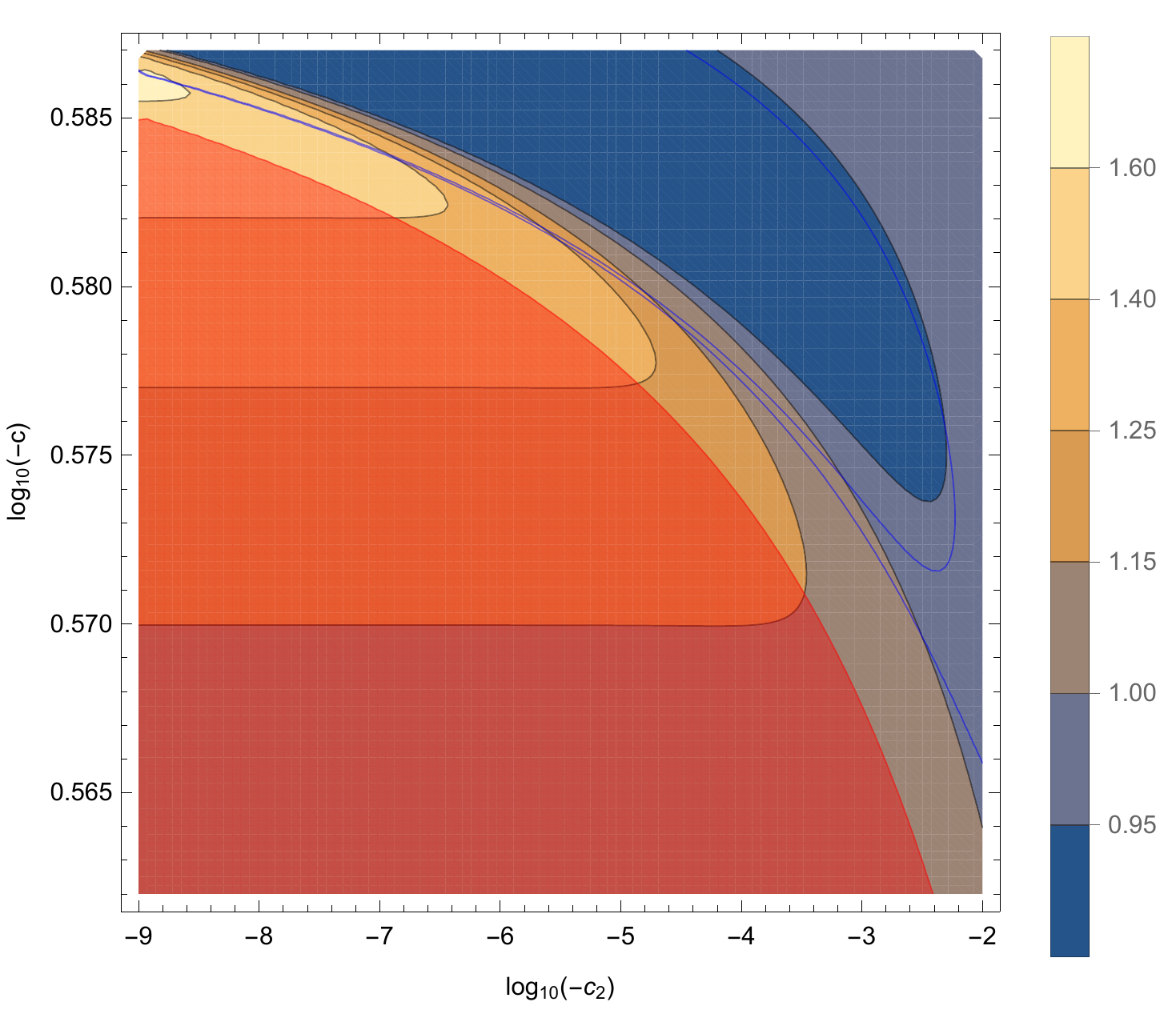}
  \caption{Same as in Fig. \ref{fig:ns_RMI3}, for the RMI4 regime with $\phi_0 = 0.1 \Mpl$ (left) and $\phi_0 = 1 \Mpl$ (right).}\label{fig:ns_RMI4}
\end{figure}

\begin{figure}
  \centering
  \includegraphics[width=9cm]{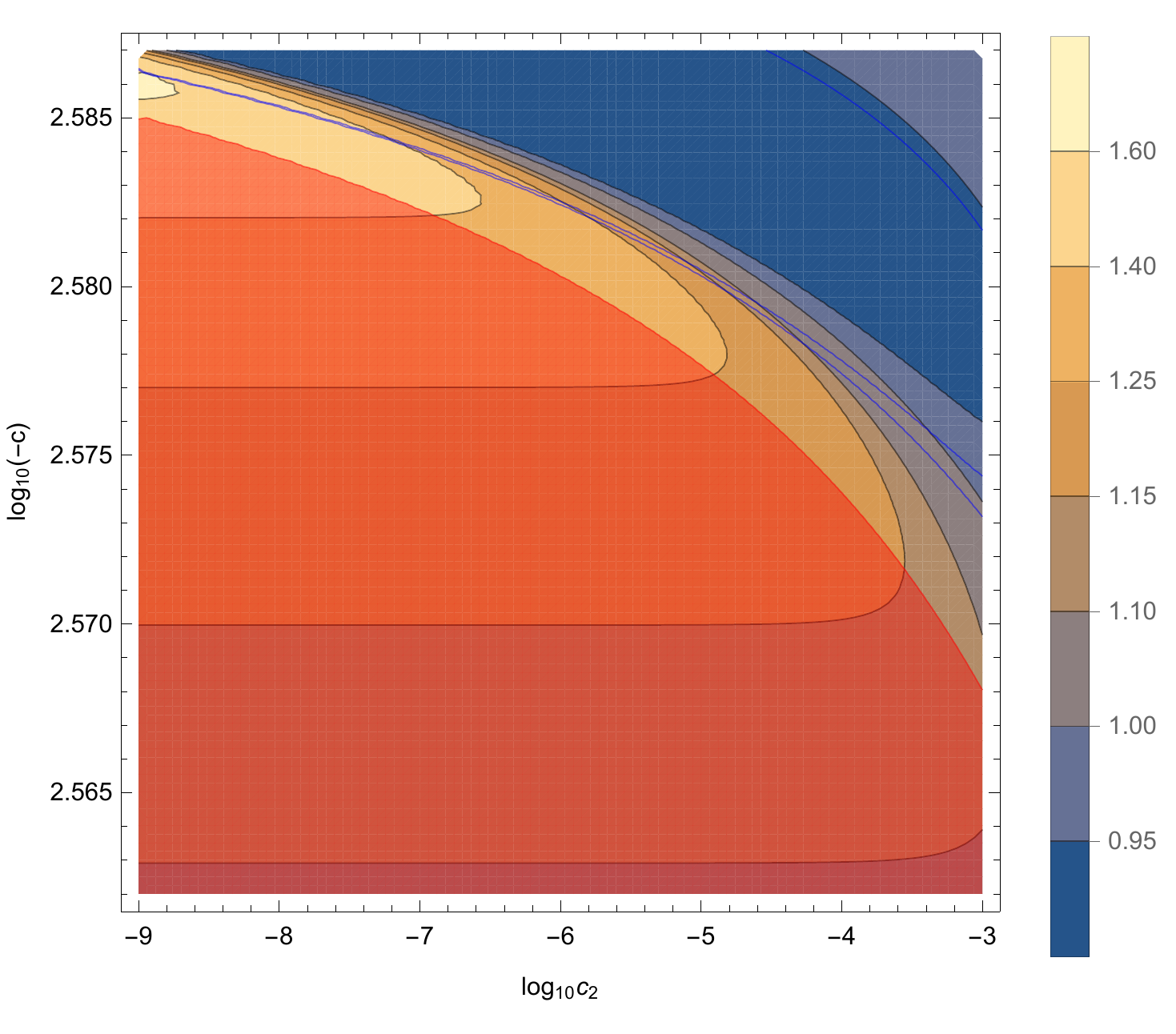}~~\includegraphics[width=9cm]{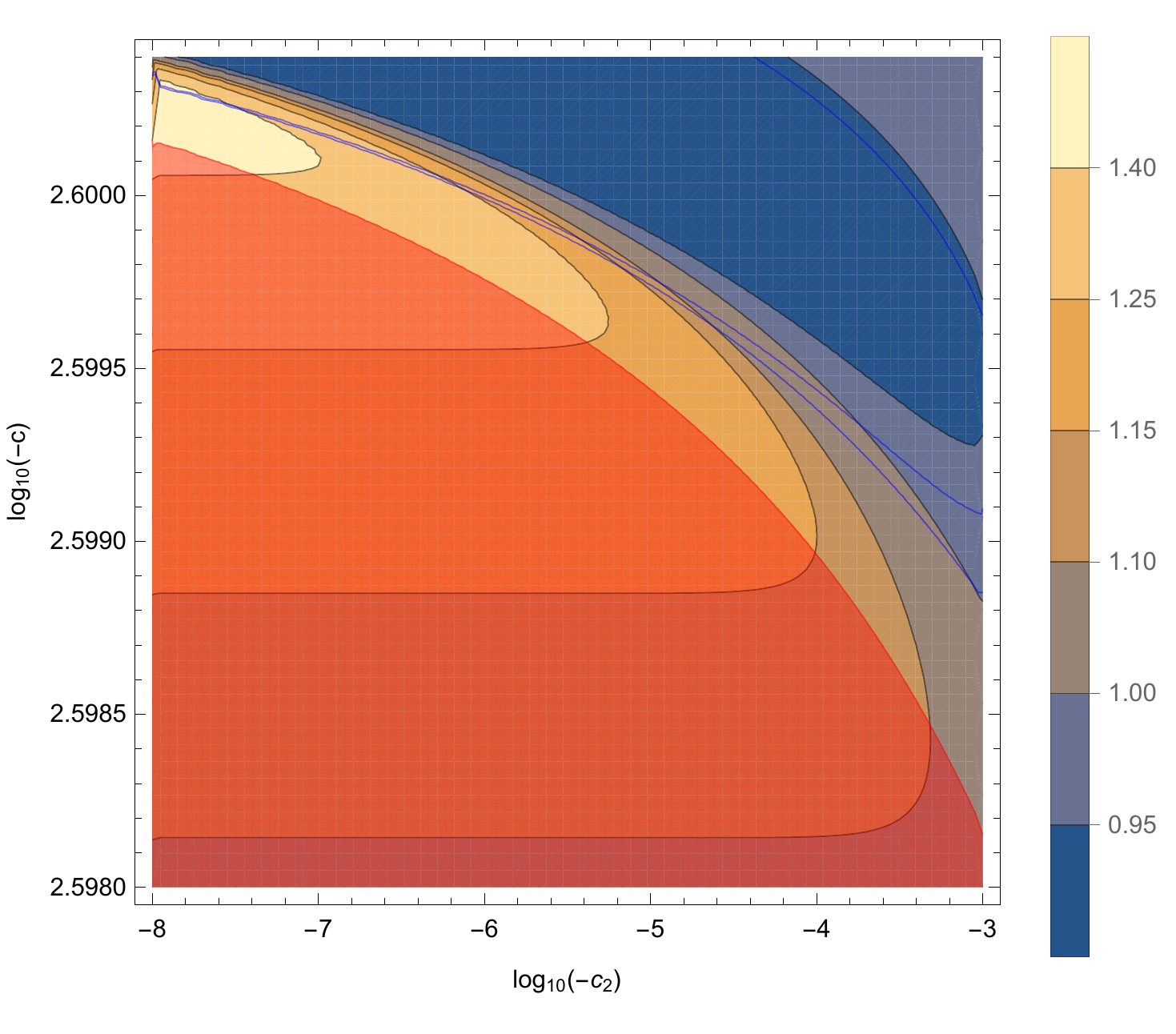}
  \caption{Same as in Fig. \ref{fig:ns_RMI3}, for the RMI3 (left) and RMI4 (right) regimes with $\phi_0 = 0.1 \Mpl$ and $M = 10^{-4} \sqrt{\Lambda} / \Mpl$.}\label{fig:ns_RMI34}
\end{figure}

We combine the discussion of RMI3 and RMI4 together, because we can deal them in the same way and the results are similar. We define the critical tachyonic point as $\phi_c = (1 - c_2) \phi_0$. When $0< c_2 <1$, it corresponds to the RMI3 regime, whereas the RMI4 regime satisfies $c_2 < 0$. Contours of the spectral index at the scale $k_d$ for the RMI3 and RMI4 regimes are presented in Figs. \ref{fig:ns_RMI3} and \ref{fig:ns_RMI4}, respectively. Blue solid lines represent the PLANCK spectral index $2 \sigma$ confidence levels at the scale $k_p$. For both the RMI3 and RMI4 regimes, regions consistent with the PLANCK observations on scalar spectrum overlap with regions of blue tilted spectrum at the scale $k_d$, in cases of either $\phi_0 = 0.1 \Mpl$ or $\phi_0 = 1 \Mpl$. However, the case of $\phi_0 = 1 \Mpl$ is not preferred in the supersymmetric framework, since it may destory the flatness of the potential by supergravity corrections. Comparing it with that of $\phi_0 = 0.1 \Mpl$ in Fig. \ref{fig:ns_RMI3} and \ref{fig:ns_RMI4}, one can qualitatively conclude that the increasing of $\phi_0$ would suppress the absolute value of $c$. In the case of $\phi_0 = 0.1 \Mpl$, the overlapped region correspond to $c \sim -10^{2.4}$, $c_2 < 0.01$ for the RMI3 regime, and to $c \sim -10^{2.4}$, $c_2 < -0.01$ for the RMI4 regime. However, these regions cannot satisfy the constraints on tensor modes, as they do not overlap with the red regions in Figs. \ref{fig:ns_RMI3} and \ref{fig:ns_RMI4}. We also investigate the case of the coupling $M = 10^{-4} \sqrt{\Lambda} / \Mpl$, and obtain similar results as shown in Fig. \ref{fig:ns_RMI34}. Compared with the $M = 0.01 \sqrt{\Lambda} / \Mpl$ case above and the minimally coupled case in Ref. \cite{Clesse:2014pna}, we note that for the potentially interesting region the kinetic coupling lifts up the absolute value of the parameter $c$, and make the critical point $\phic$ closer to the minimum point of the potential.

\section{Conclusion} \label{sec:ccl}
We have investigated the observational prospects of CMB spectral distortions generated by non-minimal derivative-coupled inflation in a model-oriented approach. In order to reveal the effects of kinetic coupling clearly, we consider the situations in the high friction limit $M^2 \ll H^2$. After examining all 49 models listed in Ref. \cite{Martin2013a}, there are only $5$ single-field models that can lead to increasing small-scale power, which is the necessary condition for a detectable distortion signal for a future PIXIE-like experiment. These models are hybrid inflation, non-canonical K\"{a}hler inflation, generalized MSSM inflation, generalized renormalization inflection point inflation, and running-mass inflation. For simultaneous consistency with PLANCK observations of the primordial scalar power spectrum and giving rise to a blue tilted spectrum on distortion scales, the potential regimes have been identified for all these models. For all models, the corresponding regions of parameter space are strongly tuned. Compared with the minimally coupled models in Ref. \cite{Clesse:2014pna}, we note that the kinetic coupling suppresses model parameters to be sub-Planckian for the VHI, GMSSMI and GRIPI models. It helps the models to agree with theoretical considerations from the supersymmetric framework. For example, when $M = 10^{-5} \sqrt{\Lambda} / \Mpl$, the value of parameter $\phi_0$ for the GMSSMI model can drop off to the order of $10^{-4} \Mpl$, which is required by the MSSM scenario \cite{Allahverdi2006}. For the NCKI model, it can lead to a small scale enhanced spectrum only if $\beta \gg \alpha$, where the model becomes an approximate VHI model. For the RMI model, there are two inflationary regimes (RMI3 and RMI4) potentially detected by a PIXIE-like experiment. In the case of $\phi_0 = 0.1 \Mpl$, the model parameter $c$ needs to be of the order of $10^{2}$ to induce a blue spectrum on small scales. In the original model \cite{Covi2003,Covi2004}, the absolute value of $c$ should be of the order $10^{-1}$ to $10^{-2}$. A larger value would not allow inflation to take place. However, with kinetic coupling, the slow-roll parameters are suppressed by a factor $M^2 / 3 H^2$, such that the slow-roll conditions can be still preserved when $|c|$ is large.

Nevertheless, all the potential regimes have been ruled out by the constraints on tensor modes at $95\%$ CL. For these potential regimes, the corresponding values of the tensor-to-scalar ratio exceed the upper limit $r_{0.002} < 0.065$ given by the observations \cite{Array:2015xqh,Aghanim2018eyx,Akrami2018}. We have checked different cases by adjusting the strength of the kinetic coupling for each model. For VHI, GMSSMI, and GRIPI, the model parameters $\mu_{\rr VHI}$ and $\phi_0$ change proportionally with the coupling constant $M$, such that behaviors of slow-roll parameters, spectral index, and tensor-to-scalar ratio are very similar for different values of $M$. For the RMI3 and RMI4 regimes, the increase of $M$ makes the critical point $\phi_c$ closer to the minimum potential point $\phi = \phi_0$. But the behaviors of the spectral index and the tensor-to-scalar ratio are still similar. Thus, there is no kinetically coupled single field inflation model that can give rise to any observable CMB distortion for a PIXIE-like experiment.

\section*{Acknowledgements}

The authors warmly thank Bj\"{o}rn Garbrecht, S\'{e}bastien Clesse, Yungui Gong, and Qing Gao for useful discussions and comments. The authors are also very grateful to the reviewer for detailed comments and suggestions that have helped improve the manuscript substantially. RD and YZ received support from the National Natural Science Foundation of China (Grant No. 11705132) and the Fundamental Research Funds for the Central Universities (WUT:2018IVB018).

\bibliography{cmbdist}

\end{document}